\documentclass[english]{IEEEtran}
\usepackage[T1]{fontenc}
\usepackage[latin9]{inputenc}
\usepackage{color}
\usepackage{verbatim}
\usepackage{amsthm}
\usepackage{amsmath}
\usepackage{amssymb}
\usepackage{graphicx}

\makeatletter

\providecommand{\tabularnewline}{\\}

  \theoremstyle{plain}
  \newtheorem{prop}{\protect\propositionname}
  \theoremstyle{definition}
  \newtheorem{example}{\protect\examplename}
  \theoremstyle{plain}
  \newtheorem{thm}{\protect\theoremname}
  \theoremstyle{plain}
  \newtheorem{lem}{\protect\lemmaname}
  \theoremstyle{remark}
  \newtheorem{rem}{\protect\remarkname}

\newtheorem{assumption}{Assumption}
\usepackage{subfig}

\makeatother

\usepackage{babel}
\providecommand{\examplename}{Example}
\providecommand{\lemmaname}{Lemma}
\providecommand{\propositionname}{Proposition}
\providecommand{\remarkname}{Remark}
\providecommand{\theoremname}{Theorem}

\begin{document}

\title{Mixed-Timescale Precoding and Cache Control in Cached MIMO Interference
Network}

\author{{\normalsize{An Liu, }}\textit{\normalsize{Member IEEE}}{\normalsize{,
and Vincent Lau,}}\textit{\normalsize{ Fellow IEEE}}{\normalsize{,\\Department
of Electronic and Computer Engineering, Hong Kong University of Science
and Technology}}%

}
\maketitle
\begin{abstract}
Consider media streaming in MIMO interference networks whereby multiple
base stations (BS) simultaneously deliver media to their associated
users using fixed data rates. The performance is fundamentally limited
by the cross-link interference. We propose a \textit{cache-induced
opportunistic} \textit{cooperative MIMO} (CoMP) for interference mitigation.
By caching a portion of the media files, the BSs opportunistically
employ CoMP to transform the cross-link interference into spatial
multiplexing gain. We study a mixed-timescale optimization of MIMO
precoding and cache control to minimize the transmit power under the
rate constraint. The cache control is to create more CoMP opportunities
and is adaptive to the long-term popularity of the media files. The
precoding is to guarantee the rate requirement and is adaptive to
the channel state information and \textit{cache state} at the BSs.
The joint stochastic optimization problem is decomposed into a \textit{short-term
precoding }and a \textit{long-term cache control problem}. We propose
a precoding algorithm which converges to a stationary point of the
short-term problem. Based on this, we exploit the hidden convexity
of the long-term problem and propose a low complexity and robust solution
using stochastic subgradient. The solution has significant gains over
various baselines and does not require explicit knowledge of the media
popularity.\end{abstract}

\section{Introduction}

Media streaming is going to be one of the major applications in wireless
networks. For example, it is envisioned that a significant portion
of the capacity demand in future wireless systems will come from media
streaming applications. In this paper, we consider media streaming
in MIMO interference networks whereby multiple BSs simultaneously
deliver media to their associated users using fixed data rates. The
performance of this system is fundamentally limited by the inter-cell
interference from the cross-links. In traditional cellular networks,
the inter-cell interference is mitigated using frequency planing techniques
such as frequency reuse or fractional frequency reuse \cite{ghaffar2010fractional}.
To further improve the spectrum efficiency, more advanced techniques
such as cooperative MIMO (CoMP) \cite{somekh2009cooperative} and
coordinated MIMO \cite{Foschini_POC06_CordMIMO} have been proposed
for future wireless systems. The CoMP technique can transform the
cross-link interference into spatial multiplexing gain by sharing
both real-time channel state information (CSI) and payload data among
the concerned BSs. However, it requires high capacity backhaul for
payload exchange between BSs, which is a cost bottleneck especially
in dense small cell networks. On the other hand, the coordinated MIMO
is a more cost effective technique as it only requires the exchange
of real-time CSIs among the BSs to perform joint precoding. Many MIMO
precoding optimization algorithms have been proposed for coordinated
MIMO. For example, in \cite{Luo_TSP11_WMMSE}, a WMMSE algorithm is
proposed to find a stationary point of the weighted sum-rate maximization
problem for multi-cell downlink systems. In \cite{Liu_IT10s_Duality_BMAC,Liu_10sTSP_Fairness_rate_polit_WF,Liu_10sTSP_MLC},
the authors proposed \textit{polite water-filling} method for precoding
optimization in B-MAC interference networks based on the duality principle
of interference networks. Although the coordinated MIMO requires smaller
backhaul capacity, the overall performance is usually much lower than
that of CoMP. Recently, there have been some works conducted on multi-cell
coordination with consideration of backhaul limitation. In \cite{Huang_TSP11_DMBF},
a distributed and hierarchical solution of joint beamforming and power
allocation was proposed to maximize the worst-user SINR in time-division-duplex
(TDD) multicell downlink systems where only limited inter-cell information
exchange is permitted. In \cite{Rao_TWC13_McellICIC}, random matrix
theory is leveraged to design a distributed joint beamforming and
power control algorithm that only requires statistical information.
Such design reduces the amount of control signaling over the backhaul. 

An interesting question is, can we achieve the CoMP gain with reduced
backhaul bandwidth consumption? We show that this is possible for
media streaming applications by using a novel \textit{cache-induced
opportunistic CoMP} scheme proposed in this paper. Specifically, we
can opportunistically transform the interference network into a CoMP
broadcast channel by caching a portion of the media files at the BSs.
As a result, there are two transmission modes at the physical layer,
namely, the \textit{CoMP mode} and the \textit{coordinated MIMO mode},
depending on the cache state at the BSs. If the payload data accessed
by each user exists in the cache of the BSs, the BSs can engage in
CoMP and therefore, enjoy a large performance gain without consuming
the backhaul bandwidth. Otherwise, coordinated MIMO is employed at
the BSs to serve the users. Hence, there is a cache-induced topology
change in the physical layer (dynamic CoMP opportunity) of the MIMO
interference network. As such, a MIMO interference network employing
the cache-induced opportunistic CoMP is called a \textit{cached MIMO
interference network} in this paper. With high capacity caches at
the BSs and a proper caching strategy, the opportunity of CoMP in
the cached MIMO interference network can be very large and thus the
proposed solution will have a significant gain over the coordinated
MIMO scheme with even smaller backhaul consumption. Note that in the
proposed solution, the reduced backhaul consumption is due to the
reduced payload data transmission over the backhaul. The payload data
transmission consumes much more backhaul bandwidth than the exchange
of control signaling because the former needs to be done on a per-symbol
basis but the latter needs to be done on a per frame basis. Hence
the backhaul saving of the proposed solution is much more significant
compared to those only reduce the control signaling in the backhaul
\cite{Huang_TSP11_DMBF,Rao_TWC13_McellICIC}. Since the cost of hard
disks is much lower than the cost of optical fiber backhaul, the proposed
solution is very cost effective. 

The performance of the proposed solution depends heavily on the dynamic
caching strategy (which affects the opportunity of CoMP) and the MIMO
precoding design. We study a mixed-timescale joint optimization of
MIMO precoding and cache control in cached MIMO interference networks
to minimize the average sum transmit power subject to fixed data rate
constraints for all users. The role of cache control is to create
more CoMP opportunities and is adaptive to long-term popularity of
the media files (long-term control). The role of MIMO precoding optimization
is to exploit the CoMP opportunities (induced by the cache) to guarantee
the individual rate constraints for each user. As such, it is adaptive
to the instantaneous CSI and the \textit{cache state} at the BSs.
There are several first order technical challenges to be addressed.
\begin{itemize}
\item \textbf{Limited Cache Size}: The performance gain of the proposed
scheme depends heavily on the CoMP opportunity, which in turn depends
on the cache size and cache strategy. The BSs usually do not have
enough cache to store all the media files. As will be shown in Example
\ref{Naive-cahce-scheme}, when brute force caching is used, even
if a significant portion of the media files are cached at BSs, the
CoMP opportunity can still be very small and this is highly undesirable. 
\item \textbf{Non-Convex Stochastic Optimization}: The mixed-timescale joint
optimization of MIMO precoding and cache control is a non-convex stochastic
optimization problem and the complexity of finding the optimal solution
is extremely high. For example, the short-term MIMO precoding optimization
in the interference networks is well known to be a difficult non-convex
problem. Furthermore, the objective function for long-term cache control
has no closed form expression because the short-term precoding problem
has no closed form solution and the popularity of the media files
is in general unknown.
\item \textbf{Complex Coupling between Cache Control and Precoding} \textbf{Optimization}:
Caching has been widely used in fixed line P2P systems \cite{Kozat_TOM09_P2P}
and content distribution networks (CDNs) \cite{Shen_TOM04_CDN}. In
\cite{Caire_INFOCOM12_femtocache}, a FemtoCaching scheme has also
been proposed for wireless systems. However, these schemes do not
consider cache-induced opportunistic CoMP among the BSs. Hence, the
cache control in the above works is independent of the physical layer
and is fundamentally different from our case where the cache control
and physical layer are coupled together. In our case, the cache control
will affect the physical layer dynamics seen by\textbf{ }precoding\textbf{
}optimization due to different CoMP opportunities. On the other hand,
the short-term precoding strategy adopted in the physical layer will
also affect the cache control due to a different cost-reward dynamic.
\end{itemize}

To address the above challenges, we first propose a novel cache data
structure called \textit{MDS-coded random cache} which can significantly
improve the probability of CoMP. We then exploit the timescale separations
of the optimization variables to decompose the stochastic optimization
problem into a \textit{short-term precoding problem} and a \textit{long-term
stochastic cache control problem}. We generalize the WMMSE approach
in \cite{Luo_TSP11_WMMSE} to find a stationary point for the short-term
precoding problem. To solve the long-term cache control problem, we
first show that despite the non-convexity in the short-term precoding
problem, there is a hidden convexity in the long-term stochastic cache
control problem. We propose a stochastic-subgradient-like iterative
solution and show that it converges to the optimal solution of this
long-term stochastic optimization problem. The proposed solution has
low complexity and does not require explicit knowledge of the popularity
of the media files. Finally, we illustrate with simulations that the
proposed solution achieves significant gain over various baselines
under the consideration of overhead in the backhaul.

\textit{Notation}\emph{s}: The superscript $\left(\cdot\right)^{\dagger}$
denotes Hermitian. The notation $1\left(\cdot\right)$ denote the
indication function such that $1\left(E\right)=1$ if the event $E$
is true and $1\left(E\right)=0$ otherwise. The notation $\left[\mathbf{A}\right]_{i,j}$
represents the element at the $i$-th row and $j$-th column of a
matrix $\mathbf{A}$. For a square matrix $\mathbf{A}$, $\left|\mathbf{A}\right|$
denotes the determinant of $\mathbf{A}$ and $\mathbf{A}\succeq\mathbf{0}$
means that $\mathbf{A}$ is positive semidefinite. The notation $\left[a_{k}\right]_{k=1,...,K}$
denote a $K\times1$ vector whose $k$-th element is $a_{k}$.

\section{System Model\label{sec:System-Model}}

In this section, we introduce the architecture of the cached MIMO
interference networks, the physical layer (opportunistic CoMP) and
the MDS-coded random cache scheme that supports opportunistic CoMP.

\subsection{System Architecture of Cached MIMO Interference Networks}

\begin{figure}
\begin{centering}
\textsf{\includegraphics[clip,width=85mm]{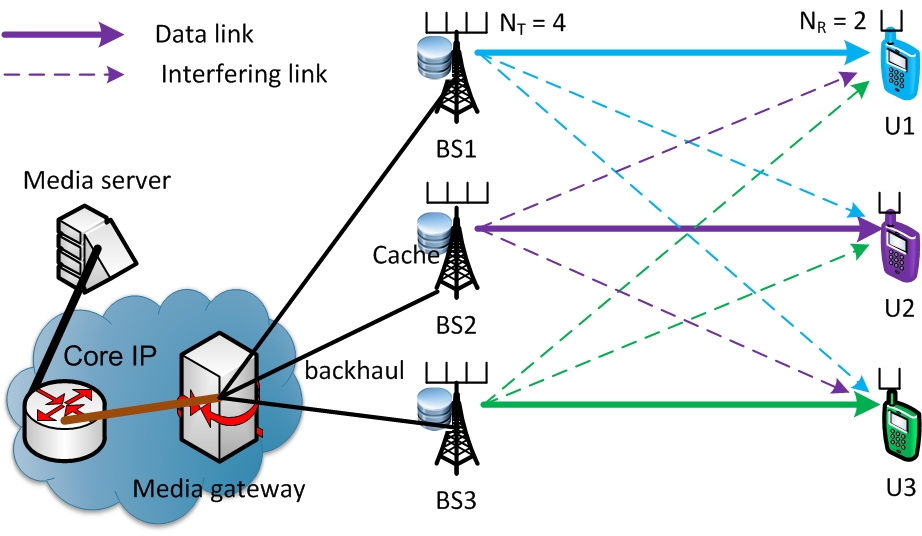}}
\par\end{centering}

\caption{\label{fig:system_model}{\small{System architecture of cached MIMO
interference network.}} }
\end{figure}

The architecture of the cached MIMO interference network is illustrated
in Fig. \ref{fig:system_model}. There are $L$ media files on the
media server. The size of the $l$-th media file is $F_{l}$ bits
and the streaming rate is denoted by $\mu_{l}$ (bits/s). There are
$K$ users streaming media files from the media server via a radio
access network (RAN) consisting of $K$ BSs and each BS is associated
with one user%
\footnote{For clarity, we consider the case where each BS is only associated
with one user. However, the solution framework can be easily extended
to the case with multiple users per BS.%
}. Each BS is equipped with $N_{T}\geq2$ antennas and each user is
equipped with $N_{R}$ antennas. The index of the media file requested
by the $k$-th user is denoted by $\pi_{k}$. Define $\pi=\left\{ \pi_{1},...,\pi_{K}\right\} $
as the user request profile (URP). We have the following assumption
on URP.

\begin{assumption}[URP Assumption]\label{asm:URP}The URP $\pi\left(t\right)$
is a slow ergodic random process (i.e., $\pi\left(t\right)$ remains
constant for a large number of time slots) according to a general
distribution.

\end{assumption}

The media packets (payload data) requested by the $k$-th user are
delivered to the $k$-th BS from the media gateway via backhaul as
illustrated in Fig. \ref{fig:system_model}. Moreover, each BS is
equipped with a cache of size $B_{C}$ bits. In this paper, the time
is partitioned into time slots indexed by $t$ with duration $\tau$.

\subsection{Benefits of Caching in MIMO Interference Networks\label{sub:Cache-enabled-Opportunistic-CoMP}}

\begin{figure}
\begin{centering}
\textsf{\includegraphics[clip,width=85mm]{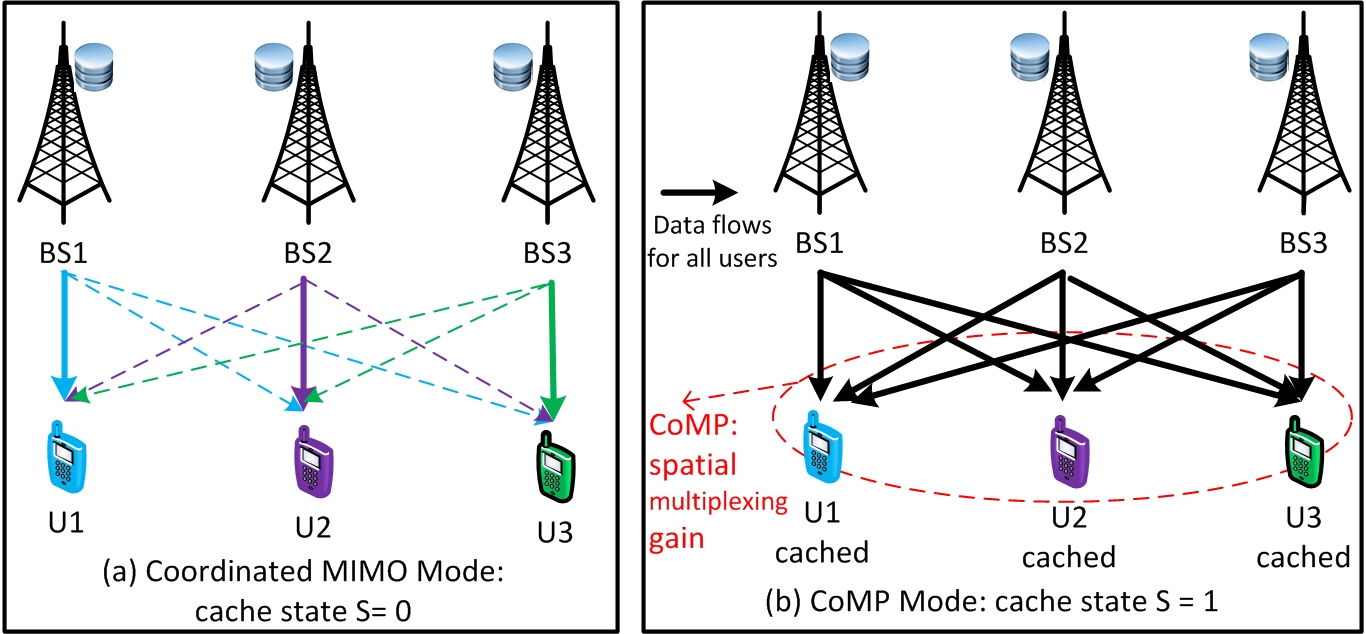}}
\par\end{centering}

\caption{\label{fig:CocaMIMO}{\small{Illustration of cache-induced opportunistic
CoMP for a RAN with $K=3$. }}}
\end{figure}

The RAN is the performance bottleneck of the system. Without caching
at each BS, the RAN forms a MIMO interference network and the performance
is limited by the inter-cell interference between the BSs. In this
section, we propose a cache-induced opportunistic CoMP which can opportunistically
use the cached media packets at the BSs to transform an interference
network into a CoMP broadcast channel as illustrated in Fig. \ref{fig:CocaMIMO}-(b).
The impact of caching at BS on the physical layer is summarized by
the \textit{cache state} defined as $S\in\left\{ 0,1\right\} $, where
$S=1$ means that $\forall k$, the current payload data requested
by user $k$ is in the cache of all the $K$ BSs and thus it is possible
for the BSs to cooperatively transmit the payload data to the users;
and $S=0$ means that the $k$-th user can only be served by the $k$-th
BS. Hence, there are two transmission modes depending on the cache
state S, namely the CoMP mode ($S=1$) and the coordinated MIMO mode
($S=0$). Fig. \ref{fig:CocaMIMO} illustrates two examples of the
data flows under different cache states $S$ (or transmission modes).
As illustrated in Fig. \ref{fig:CocaMIMO}, when $S=1$, the RAN can
enjoy significant spatial multiplexing gain \cite{Zheng_IT02_Multiplexing_tradeoff}
due to cache-induced CoMP transmission and the gain is achieved without
expensive backhauls%
\footnote{There are controversial conclusions regarding whether the conventional
CoMP is good or bad in practical cellular networks when CSI signaling
latency and payload sharing overhead in the backhaul are taken into
account. However, the degradation of CoMP performance due to CSI signaling
latency in the backhaul is not a fundamental limitation but rather
it is a limitation due to current technology. On the other hand, the
payload sharing overhead in the backhaul is a fundamental limitation
in the conventional CoMP. As such, it is the focus of the paper to
exploit the BS caching to fundamentally solve this payload backhaul
overhead in CoMP. %
}. 

We consider an OFDM based system where the wireless link between each
BS and user consists of $M\geq K$ orthogonal subcarriers. Let $\mathbf{H}_{m,k,n}\in\mathbb{C}^{N_{R}\times N_{T}}$
denote the channel matrix between user $k$ and BS $n$ on subcarrier
$m$. We have the following assumption on the global CSI $\mathbf{H}\triangleq\left\{ \mathbf{H}_{m,k,n}\right\} $.

\begin{assumption}[Channel Assumption]\label{asm:forY}$\mathbf{H}_{m,k,n}\left(t\right)$
remains constant within a time slot but is i.i.d. w.r.t. time slot
index $t$. Specifically, $\mathbf{H}_{m,k,n}\left(t\right)$ has
i.i.d. complex Gaussian entries of zero mean and variance $g_{k,n}$.

\end{assumption}

The variance $g_{k,n}$ is usually used to model the path gain between
BS $n$ and user $k$. Note that we do not require that the channel
$\mathbf{H}_{m,k,n}\left(t\right)$ is i.i.d. w.r.t. subcarrier index
$m$. We consider a centralized optimization scheme in which a central
node computes all the control variables and then transmits them to
the BSs. We assume that the central node has the knowledge of the
global CSI $\mathbf{H}$.

Then we elaborate the two transmission modes in the proposed cache-induced
opportunistic CoMP.

\textbf{Coordinated MIMO Mode}: If $S=0$, the $k$-th user can only
be served by the $k$-th BS. We consider linear precoding and MMSE
receiving for inter-cell interference cancellation. The received signal
for user $k$ on subcarrier $m$ can be expressed as:
\[
\mathbf{y}_{m,k}=\mathbf{H}_{m,k,k}\mathbf{V}_{m,k}\mathbf{s}_{m,k}+\sum_{n\neq k}\mathbf{H}_{m,k,n}\mathbf{V}_{m,n}\mathbf{s}_{m,n}+\mathbf{z}_{m,k},
\]
where $\mathbf{s}_{m,k}\in\mathbb{C}^{d_{m,k}}\sim\mathcal{CN}\left(0,\mathbf{I}\right)$
and $d_{m,k}$ are respectively the data vector and the number of
data streams for user $k$ on subcarrier $m$; $\mathbf{V}_{m,k}\in\mathbb{C}^{N_{T}\times d_{m,k}}$
is the precoding matrix for user $k$ on subcarrier $m$; and $\mathbf{z}_{m,k}\in\mathbb{C}^{N_{R}}\sim\mathcal{CN}\left(0,\mathbf{I}\right)$
is the AWGN noise vector. The MMSE receiver for user $k$ on subcarrier
$m$ is given by{\small{
\begin{equation}
\mathbf{U}_{m,k}=\left(\mathbf{\Omega}_{m,k}+\mathbf{H}_{m,k,k}\mathbf{V}_{m,k}\mathbf{V}_{m,k}^{\dagger}\mathbf{H}_{m,k,k}^{\dagger}\right)^{-1}\mathbf{H}_{m,k,k}\mathbf{V}_{m,k},\label{eq:MMSEu}
\end{equation}
}}where $\mathbf{\Omega}_{m,k}=\mathbf{I}+\sum_{n\neq k}\mathbf{H}_{m,k,n}\mathbf{V}_{m,n}\mathbf{V}_{m,n}^{\dagger}\mathbf{H}_{m,k,n}^{\dagger}$
is the interference-plus-noise covariance matrix of user $k$ on subcarrier
$m$. Then for given CSI $\mathbf{H}$, cache state $S=0$ and precoding
matrices $\mathbf{V}=\left\{ \mathbf{V}_{m,k}:\forall m,k\right\} $,
the data rate (bps) of user $k$ is
\begin{equation}
R_{k}\left(\mathbf{H},\mathbf{V}\right)=\frac{B_{W}}{M\ln2}\sum_{m=1}^{M}r_{m,k},\label{eq:Rk}
\end{equation}
where $B_{W}$ is the bandwidth of the system; and
\[
r_{m,k}=\textrm{log}\left|\mathbf{I}+\mathbf{H}_{m,k,k}\mathbf{V}_{m,k}\mathbf{V}_{m,k}^{\dagger}\mathbf{H}_{m,k,k}^{\dagger}\mathbf{\Omega}_{m,k}^{-1}\right|.
\]
The corresponding sum transmit power is
\begin{equation}
P\left(\mathbf{V}\right)=\sum_{k=1}^{K}\sum_{m=1}^{M}\textrm{Tr}\left(\mathbf{V}_{m,k}\mathbf{V}_{m,k}^{\dagger}\right).\label{eq:Pow}
\end{equation}

\textbf{CoMP Mode}: If $S=1$, the $K$ users are served using CoMP
between the BSs. Similarly, we consider linear precoding and MMSE
receiving for interference cancellation. The received signal for user
$k$ on subcarrier $m$ can be expressed as:
\[
\widetilde{\mathbf{y}}_{m,k}=\widetilde{\mathbf{H}}_{m,k}\widetilde{\mathbf{V}}_{m,k}\widetilde{\mathbf{s}}_{m,k}+\sum_{n\neq k}\widetilde{\mathbf{H}}_{m,k}\widetilde{\mathbf{V}}_{m,n}\widetilde{\mathbf{s}}_{m,n}+\mathbf{z}_{m,k},
\]
where $\widetilde{\mathbf{H}}_{m,k}=\left[\mathbf{H}_{m,k,1},...,\mathbf{H}_{m,k,K}\right]\in\mathbb{C}^{N_{R}\times KN_{T}}$
is the composite channel matrix between all the BSs and user $k$;
$\widetilde{\mathbf{s}}_{m,k}\in\mathbb{C}^{\widetilde{d}_{m,k}}\sim\mathcal{CN}\left(0,\mathbf{I}\right)$
and $\widetilde{d}_{m,k}$ are respectively the data vector and the
number of data streams for user $k$ on subcarrier $m$; and $\widetilde{\mathbf{V}}_{m,k}\in\mathbb{C}^{KN_{T}\times\widetilde{d}_{m,k}}$
is the composite precoding matrix for user $k$ on subcarrier $m$.
In CoMP mode, the MMSE receiver for user $k$ on subcarrier $m$ is
given by
\[
\widetilde{\mathbf{U}}_{m,k}=\left(\widetilde{\mathbf{\Omega}}_{m,k}+\widetilde{\mathbf{H}}_{m,k}\widetilde{\mathbf{V}}_{m,k}\widetilde{\mathbf{V}}_{m,k}^{\dagger}\widetilde{\mathbf{H}}_{m,k}^{\dagger}\right)^{-1}\widetilde{\mathbf{H}}_{m,k}\widetilde{\mathbf{V}}_{m,k},
\]
where $\widetilde{\mathbf{\Omega}}_{m,k}=\mathbf{I}+\sum_{n\neq k}\widetilde{\mathbf{H}}_{m,k}\widetilde{\mathbf{V}}_{m,n}\widetilde{\mathbf{V}}_{m,n}^{\dagger}\widetilde{\mathbf{H}}_{m,k}^{\dagger}$
is the interference-plus-noise covariance matrix of user $k$ on subcarrier
$m$. Then for given CSI $\mathbf{H}$, cache state $S=1$ and precoding
matrices $\widetilde{\mathbf{V}}=\left\{ \widetilde{\mathbf{V}}_{m,k}:\forall m,k\right\} $,
the data rate (bps) of user $k$ is
\begin{equation}
\widetilde{R}_{k}\left(\mathbf{H},\widetilde{\mathbf{V}}\right)=\frac{B_{W}}{M\ln2}\sum_{m=1}^{M}\widetilde{r}_{m,k},\label{eq:Rk1}
\end{equation}
where 
\[
\widetilde{r}_{m,k}=\textrm{log}\left|\mathbf{I}+\widetilde{\mathbf{H}}_{m,k}\widetilde{\mathbf{V}}_{m,k}\widetilde{\mathbf{V}}_{m,k}^{\dagger}\widetilde{\mathbf{H}}_{m,k}^{\dagger}\widetilde{\mathbf{\Omega}}_{m,k}^{-1}\right|.
\]
The corresponding sum transmit power is
\begin{equation}
\widetilde{P}\left(\widetilde{\mathbf{V}}\right)=\sum_{k=1}^{K}\sum_{m=1}^{M}\textrm{Tr}\left(\widetilde{\mathbf{V}}_{m,k}\widetilde{\mathbf{V}}_{m,k}^{\dagger}\right).\label{eq:Pow1}
\end{equation}

The choice of the number of data streams $\left\{ d_{m,k}\right\} $
and $\left\{ \widetilde{d}_{m,k}\right\} $ is an important problem.
In \cite{Luo_TSP11_WMMSE}, the number of data streams is treated
as a system parameter and there is no discussion about how to choose
this parameter. In this paper, we show that it will not lose ``optimality''
to choose the number of data streams for the coordinated MIMO mode
to be $d_{m,k}=d\triangleq\textrm{min}\left(N_{T},N_{R}\right),\forall m,k$,
in the sense that for any set of precoding matrices $\mathbf{V}^{'}$
that achieves a rate point with certain transmit power at each BS,
there exists a set of precoding matrices $\mathbf{V}=\left\{ \mathbf{V}_{m,k}\in\mathbb{C}^{N_{T}\times d}:\forall m,k\right\} $
with $d=\textrm{min}\left(N_{T},N_{R}\right)$ such that an equal
or larger rate point can be achieved with equal or less transmit power
at each BS. This result is formally stated in the following proposition.
\begin{prop}
[Choice of the number of data streams]\label{prop:Optimal-number-ofd}For
any set of precoding matrices $\mathbf{V}^{'}=\left\{ \mathbf{V}_{m,k}^{'}\in\mathbb{C}^{N_{T}\times d_{m,k}}:\forall m,k\right\} $
with $d_{m,k}\in\mathbb{Z}_{+},\forall m,k$, there exists a set of
precoding matrices $\mathbf{V}=\left\{ \mathbf{V}_{m,k}\in\mathbb{C}^{N_{T}\times d}:\forall m,k\right\} $
with $d=\textrm{min}\left(N_{T},N_{R}\right)$ such that
\[
R_{k}\left(\mathbf{H},\mathbf{V}\right)\geq R_{k}\left(\mathbf{H},\mathbf{V}^{'}\right),\: P_{k}\left(\mathbf{V}\right)\leq P_{k}\left(\mathbf{V}^{'}\right),\forall k,\mathbf{H},
\]
where $P_{k}\left(\mathbf{V}\right)=\sum_{m=1}^{M}\textrm{Tr}\left(\mathbf{V}_{m,k}\mathbf{V}_{m,k}^{\dagger}\right)$
is the transmit power at BS $k$.
\end{prop}

Please refer to Appendix \ref{sub:Proof-of-Proposition-Optd} for
the proof. Similarly, it will not lose ``optimality'' to choose
the number of data streams for the CoMP mode to be $\widetilde{d}_{m,k}=\widetilde{d}\triangleq\textrm{min}\left(KN_{T},N_{R}\right),\forall m,k$.

\subsection{Random Caching using Maximum Distance Separable (MDS) Code\label{sub:Outline-of-cache}}

The overall performance gain of the cache-induced opportunistic CoMP
depends heavily on the probability of $S=1$. In this section, we
propose a novel MDS-coded random cache scheme which makes the best
use of the BS cache to increase $\textrm{Pr}\left[S=1\right]$. We
first use an example to show that, with a naive cache scheme, the
CoMP opportunity ($\textrm{Pr}\left[S=1\right]$) can be very small
even if a significant portion of the media files are stored at the
BS cache.
\begin{example}
[Brute-force cache data structure]\label{Naive-cahce-scheme}Suppose
that there are $K=4$ BS-user pairs and $L=4$ media files with equal
size of $F$ bits. The $k$-th media file is requested by the $k$-th
user. Each BS randomly stores half of the media packets (i.e., $0.5F$
bits) for each media file. Then the probability that the packets requested
by a single user are in the cache of all the BSs is $0.5^{K}=0.0625$.
However, the probability of $S=1$ is only $0.0625^{K}<0.00002$.
\end{example}

Hence, a more intelligent cache scheme is needed. In the following,
we propose a novel MDS-coded random cache data structure which can
significantly improve the probability of CoMP.
\begin{figure}
\begin{centering}
\textsf{\includegraphics[clip,width=85mm]{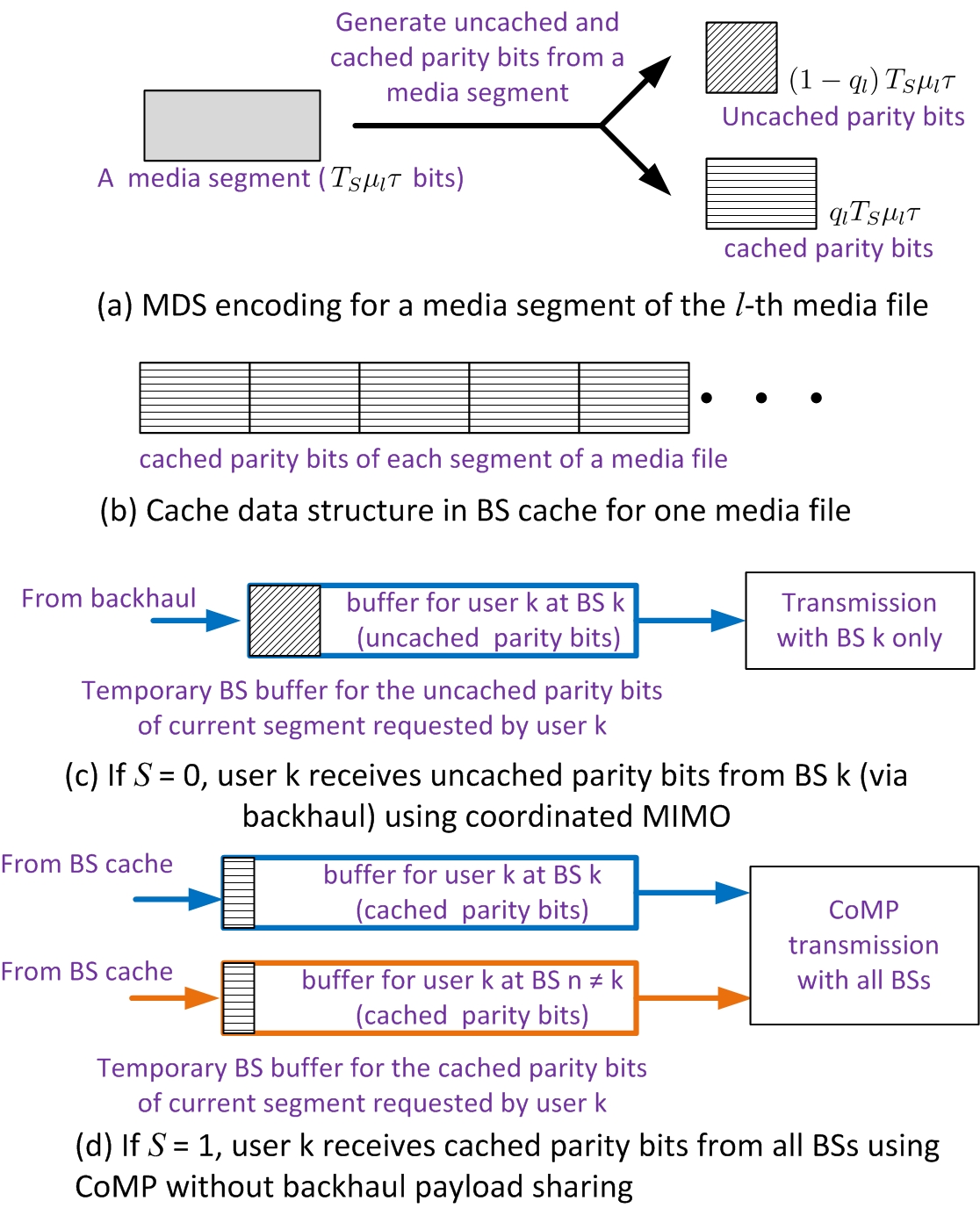}}
\par\end{centering}

\caption{\label{fig:MDS-cache}{\small{MDS-coded cache data structure and cache
usage.}}}
\end{figure}

\subsubsection*{Cache Data Structure}

Each media file is divided into segments. Specifically, each segment
of the $l$-th media file contains $T_{S}\mu_{l}\tau$ bits, where
$T_{S}\gg1$ is some integer, and it is encoded into $T_{S}\mu_{l}\tau$
parity bits using an ideal MDS rateless code as illustrated in Fig.
\ref{fig:MDS-cache}-(a). An MDS rateless code generates an arbitrarily
long sequence of parity bits from an information packet of $L_{S}$
bits ($L_{S}$ can be any positive integer), such that if the decoder
obtains any $L_{S}$ parity bits, it can recover the original $L_{S}$
information bits. In practice, the MDS rateless code can be implemented
using Raptor codes \cite{Shokrollahi_TIT06_Raptorcode} at the cost
of a small redundancy overhead. The cache at each BS stores $q_{l}T_{S}\mu_{l}\tau$
parity bits for every segment of the $l$-th media file as illustrated
in Fig. \ref{fig:MDS-cache}-(b), where $q_{l}\in\left\{ 0,\frac{1}{T_{S}},...,\frac{T_{S}-1}{T_{S}},1\right\} $
is called the \textit{cache control variable}. Such a cache data structure
is more flexible than the brute-force cache data structure in Example
\ref{Naive-cahce-scheme} in the sense that we can control when to
use the cached data.

\subsubsection*{Random Cache Usage}

\begin{figure}
\begin{centering}
\textsf{\includegraphics[clip,width=85mm]{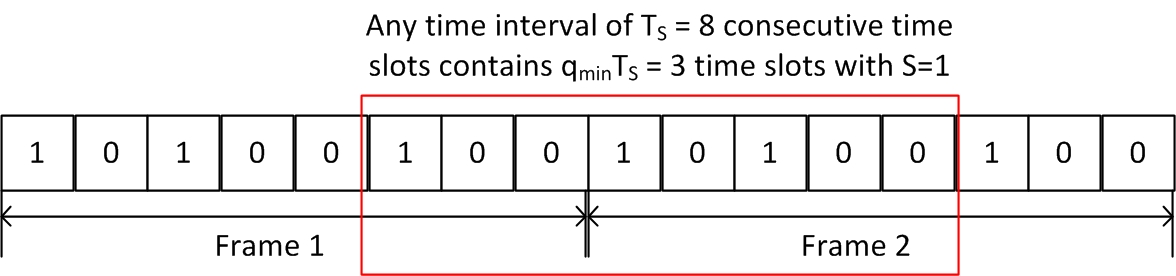}}
\par\end{centering}

\caption{\label{fig:GenS}{\small{An example on how to generate the cache state
$S$. At each time slot, the cache state $S$ is first generated by
a random cache state generator at the central node (as illustrated
in Fig. \ref{fig:moduleconn}) and then it is broadcast to the BSs.
In this example, we assume that $T_{S}=8$ and $q_{\textrm{min}}=3/8$.
The index set $\mathcal{T}_{S}=\left\{ 1,3,6\right\} $. The red box
contains a time interval of $T_{S}=8$ consecutive time slots. It
is easy to see that for any time interval of $T_{S}=8$ consecutive
time slots, there are $q_{\textrm{min}}T_{S}=3$ time slots with $S=1$.}}}
\end{figure}

The cache usage is determined by the cache state $S$ as illustrated
in Fig. \ref{fig:MDS-cache}-(c,d). Using the above MDS-coded cache
data structure, we can actively control the cache state $S$ for each
time slot. The key to increasing the probability of CoMP in the system
is to align the transmissions of the cached data for different users
as much as possible. Specifically, for given cache control vector
$\mathbf{q}=\left[q_{1},...,q_{L}\right]^{T}$ and URP $\pi$, let
$q_{\textrm{min}}=\underset{1\leq k\leq K}{\textrm{min}}\left\{ q_{\pi_{k}}\right\} $.
Then conditioned on a given $\mathbf{q}$ and $\pi$, the cache state
$S$ is generated by a random cache state generator at the central
node (as illustrated in Fig. \ref{fig:moduleconn}) using the following
method.{\small{ }}First, time is divided into frames where each frame
contains $T_{S}$ time slots. Then the central node randomly generates
an index set $\mathcal{T}_{S}\subseteq\left\{ 1,2,...,T_{S}\right\} $
such that $\left|\mathcal{T}_{S}\right|=q_{\textrm{min}}T_{S}$. Note
that for given $\mathbf{q}$ and $\pi$, $\mathcal{T}_{S}$ is only
generated for once and it remains constant until $\mathbf{q}$ and
$\pi$ changes to a new value. Suppose that the current time slot
$t$ is the $i$-th time slot in the current frame. Then if $i\in\mathcal{T}_{S}$,
we let $S\left(t\right)=1$; and otherwise, we let $S\left(t\right)=0$.
Finally, the central node broadcasts the cache state $S\left(t\right)$
to the BSs. Fig. \ref{fig:GenS} gives an example of how to generate
the cache state $S$ for each time slot. It can be seen that for any
time interval of $T_{S}$ consecutive time slots, there are $q_{\textrm{min}}T_{S}$
time slots with $S=1$ and $\left(1-q_{\textrm{min}}\right)T_{S}$
time slots with $S=0$, as illustrated in Fig. \ref{fig:GenS}.

The BSs decide when to do MIMO cooperation according to the generated
cache state $S$. If $S=1$, the BSs employ CoMP to jointly transmit
the cached parity bits to the $K$ users without consuming the backhaul
bandwidth as illustrated in Fig. \ref{fig:MDS-cache}-(d). Otherwise,
BS $k$ obtains the parity bits requested by user $k$ from the backhaul
and transmits them to user $k$ using the coordinated MIMO transmission
mode as illustrated in Fig. \ref{fig:MDS-cache}-(c).

\subsubsection*{Media Decoding at Each User}

\begin{figure}
\begin{centering}
\textsf{\includegraphics[clip,width=85mm]{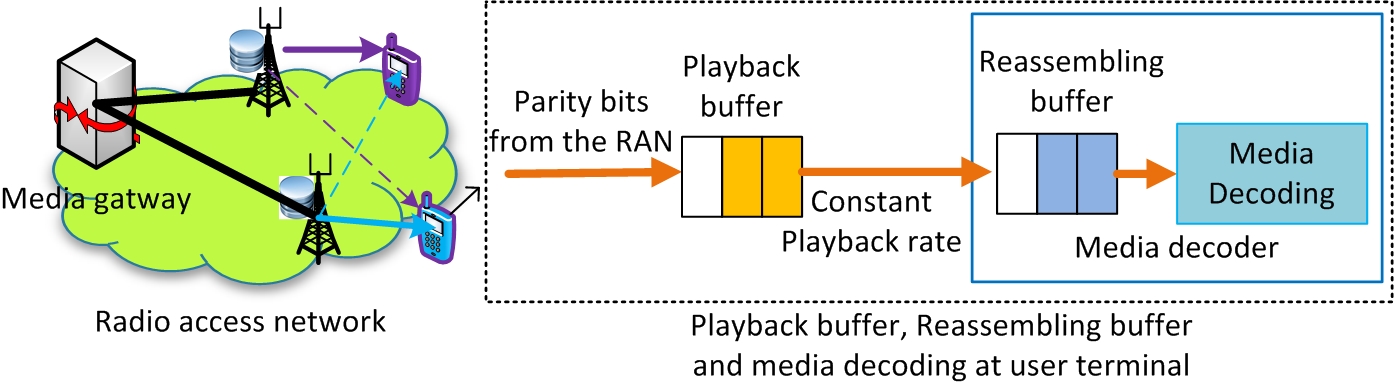}}
\par\end{centering}

\caption{\label{fig:Rasbuff}{\small{An illustration of playback buffer, reassembling
buffer and media decoding at the user. Once the total number of parity
bits for current segment at the reassembling buffer is equal to $T_{S}\mu_{l}\tau$,
the whole segment is decoded at the media decoder and the reassembling
buffer is cleared so that the media decoder can read the next segment
from the playback buffer.}}}
\end{figure}

At user $k$, the process of receiving and decoding a segment of the
$l$-th media file, where $l=\pi_{k}$, is summarized as follows.
There is a playback buffer%
\footnote{In general, the arrival packets from the RAN is burst and the playback
buffer is used to maintain a constant playback rate at the media decoder.%
} and a media decoder at the user terminal. The media decoder has a
\textit{reassembling buffer} as illustrated in Fig. \ref{fig:Rasbuff}.
At each time slot, if $S=1$, user $k$ receives $\mu_{l}\tau$ cached
parity bits of current segment from all BSs using CoMP and stores
them in the playback buffer. If $S=0$, user $k$ receives $\mu_{l}\tau$
uncached parity bits of current segment from BS $k$ using coordinated
MIMO and stores them in the playback buffer. User $k$ keeps receiving
parity bits from the RAN until the total number of received parity
bits for current segment is equal to $T_{S}\mu_{l}\tau$. Then, user
$k$ starts to receive the next segment from the RAN. On the other
hand, the media decoder keeps reading parity bits from the playback
buffer at a constant playback rate (which is equal to $\mu_{l}$)
and storing them in the \textit{reassembling buffer} until the total
number of parity bits for current segment at the reassembling buffer
is equal to $T_{S}\mu_{l}\tau$. Then the whole segment is decoded
and the reassembling buffer is cleared so that the media decoder can
read the next segment from the playback buffer. Clearly, it takes
$T_{S}$ time slots for user $k$ to receive all the $T_{S}\mu_{l}\tau$
parity bits and the number of time slots with $S=1$ is $q_{\textrm{min}}T_{S}$.
Hence, user $k$ receives a total number of $q_{\textrm{min}}T_{S}\mu_{l}\tau$
cached parity bits for each segment, which is feasible (i.e., there
is no BS cache underflow) since each BS stores $q_{l}T_{S}\mu_{l}\tau\geq q_{\textrm{min}}T_{S}\mu_{l}\tau$
parity bits for every segment of the $l$-th media file.

The following example illustrates the advantage of MDS-coded random
cache scheme.
\begin{example}
[Advantage of MDS-coded random cache]\label{Advantage-of-MDS-coded}Consider
the setup in Example \ref{Naive-cahce-scheme}. We have $q_{\pi_{k}}=0.5,\forall k$
and thus the probability of $S=1$ is $0.5$, which is much larger
than that of the brute-force caching scheme in Example 1 ($<0.00002$).
\end{example}

Compared with the brute-force caching scheme in Example \ref{Naive-cahce-scheme},
the probability of CoMP transmission ($S=1$) under the proposed MDS-coded
random cache is $\textrm{min}_{k}\left\{ q_{\pi_{k}}\right\} $ versus%
\footnote{For the brute-force caching scheme, the probability that the packets
requested by user $k$ are in the cache of all the BSs is $q_{\pi_{k}}^{K}$.
For $k\neq l$, whether the packets requested by user $k$ is in the
BS caches is independent of whether the packets requested by user
$l$ is in the BS caches. Hence, the probability of $S=1$ is \textbf{$\prod_{k=1}^{K}q_{\pi_{k}}^{K}$}.%
} $\prod_{k=1}^{K}q_{\pi_{k}}^{K}$. This represents a first order
improvement in the opportunity of CoMP gain. Yet, there is a fundamental
tradeoff between the performance gain and the BS cache size. Intuitively,
the more popular the media file is, the larger portion of its parity
bits should be stored in the BS cache to increase the CoMP probability.
Hence, the value of $\mathbf{q}$ must be carefully controlled to
achieve the best tradeoff among performance and the BS cache size.
As such, the cache control variable is parameterized by the vector
$\mathbf{q}=\left[q_{1},...,q_{L}\right]^{T}$.

\section{Mixed Timescale Precoding and Cache Control\label{sec:Optimization-Problem-Formulation}}

In this section, we formulate a mixed-timescale optimization problem
for media streaming under cache-induced opportunistic CoMP. The control
variables are partitioned into \textit{long-term} and \textit{short-term}
control variables. The long-term control variables (cache control
variables $\mathbf{q}$) are adaptive to the distribution of the URP
$\pi$ to induce CoMP opportunity. The short-term control variables
(precoding matrices $\mathbf{V},\widetilde{\mathbf{V}}$) are adaptive
to the instantaneous cache state $S$ and CSI $\mathbf{H}$ to exploit
the opportunistic CoMP gain and to guarantee the QoS requirements
of the users for given $\mathbf{q}$ and $\pi$.

\subsection{Problem Formulation}

\begin{figure}
\begin{centering}
\textsf{\includegraphics[clip,width=85mm]{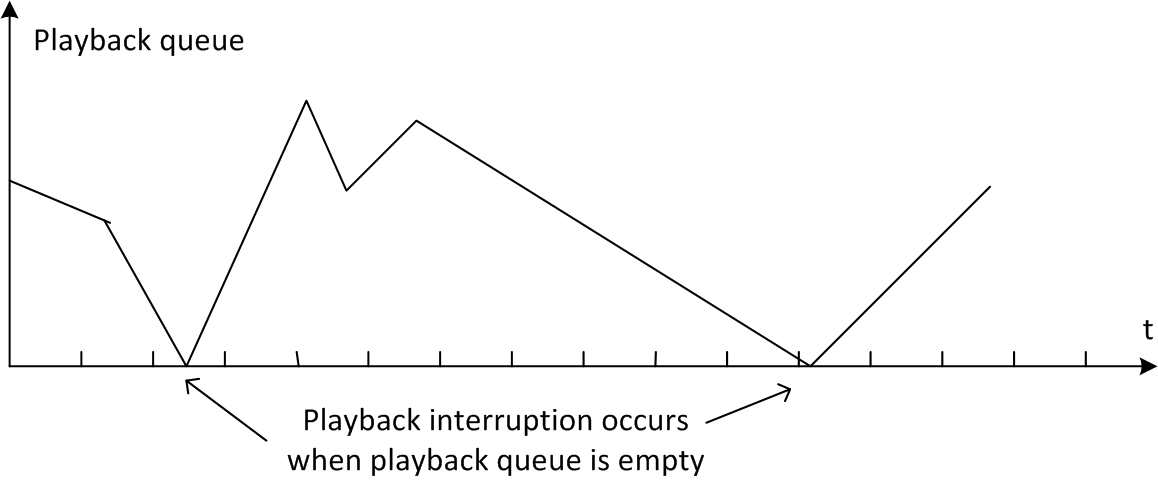}}
\par\end{centering}

\caption{\label{fig:plabackQueue}{\small{Illustration of the playback process
and playback interruption at the user.}}}
\end{figure}

For convenience, let $\mathbf{V}\left(\pi,\mathbf{H}\right)=\left\{ \mathbf{V}_{m,k}\left(\pi,\mathbf{H}\right):\forall m,k\right\} $
denote all precoding matrices under URP $\pi$, CSI $\mathbf{H}$,
and cache state $S=0$; and let $\widetilde{\mathbf{V}}\left(\pi,\mathbf{H}\right)=\left\{ \widetilde{\mathbf{V}}_{m,k}\left(\pi,\mathbf{H}\right):\forall m,k\right\} $
denote all precoding matrices under URP $\pi$, CSI $\mathbf{H}$,
and cache state $S=1$. Define $\mathcal{V}=\left\{ \mathbf{V}\left(\pi,\mathbf{H}\right),\widetilde{\mathbf{V}}\left(\pi,\mathbf{H}\right):\forall\pi,\mathbf{H}\right\} $
as the collection of precoding matrices for all possible URP, CSI,
and cache state combinations $\left\{ \pi,\mathbf{H},S\right\} $.
Then for given set of control variables $\left(\mathbf{q},\mathcal{V}\right)$
and URP $\pi$, the average sum transmit power is given by
\begin{eqnarray}
\overline{P}_{\pi}\left(\mathbf{q},\mathcal{V}\right) & = & \left(1-\textrm{min}_{k}\left\{ q_{\pi_{k}}\right\} \right)\textrm{E}\left[P\left(\mathbf{V}\left(\pi,\mathbf{H}\right)\right)|\pi\right]\nonumber \\
 &  & +\textrm{min}_{k}\left\{ q_{\pi_{k}}\right\} \textrm{E}\left[\widetilde{P}\left(\widetilde{\mathbf{V}}\left(\pi,\mathbf{H}\right)\right)|\pi\right].\label{eq:AvgPow}
\end{eqnarray}
For convenience, define the feasible sets for cache control $\mathbf{q}$,
coordinated precoding $\mathbf{V}\left(\pi,\mathbf{H}\right)$ and
CoMP precoding $\widetilde{\mathbf{V}}\left(\pi,\mathbf{H}\right)$
respectively as
\begin{eqnarray}
\mathcal{D}_{\mathbf{q}} & = & \left\{ \mathbf{q}:\: q_{l}\in\left[0,1\right],\forall l,\:\textrm{and}\:\sum_{l=1}^{L}F_{l}q_{l}\leq B_{C}\right\} ,\nonumber \\
\mathcal{D}_{\mathbf{v}}\left(\pi,\mathbf{H}\right) & = & \left\{ \mathbf{V}:\: R_{k}\left(\mathbf{H},\mathbf{V}\right)\geq\mu_{\pi_{k}},\forall k\right\} ,\nonumber \\
\mathcal{D}_{\widetilde{\mathbf{v}}}\left(\pi,\mathbf{H}\right) & = & \left\{ \widetilde{\mathbf{V}}:\:\widetilde{R}_{k}\left(\mathbf{H},\widetilde{\mathbf{V}}\right)\geq\mu_{\pi_{k}},\forall k\right\} .\label{eq:Fset}
\end{eqnarray}
In $\mathcal{D}_{\mathbf{q}}$, $\sum_{l=1}^{L}F_{l}q_{l}\leq B_{C}$
is the BS cache size constraint used to avoid BS cache overflow. Note
that $q_{l}$ is relaxed to be a real number in $\left[0,1\right]$.
This relaxation has little effect on the performance when $T_{S}\gg1$.
In $\mathcal{D}_{\mathbf{v}}\left(\pi,\mathbf{H}\right)$ ($\mathcal{D}_{\widetilde{\mathbf{v}}}\left(\pi,\mathbf{H}\right)$),
$R_{k}\left(\mathbf{H},\mathbf{V}\right)\geq\mu_{\pi_{k}}$ ($\widetilde{R}_{k}\left(\mathbf{H},\widetilde{\mathbf{V}}\right)\geq\mu_{\pi_{k}}$)
is the instantaneous rate constraint for user $k$ under URP $\pi$
and cache state $S=0$ ($S=1$). In media streaming applications,
the playback process at user $k$ can be modeled by a playback queue
with random arrival (from the BS via the RAN) and deterministic departure
as illustrated in Fig. \ref{fig:Rasbuff}. The media streaming QoS
can be represented by \textit{playback interruption probability},
which is the same as the probability of playback buffer being empty
as indicated in Fig. \ref{fig:plabackQueue}. Since the departure
process is deterministic with constant rate $\mu_{\pi_{k}}$, the
instantaneous rate constraint $R_{k}\left(\mathbf{H},\mathbf{V}\right)\geq\mu_{\pi_{k}}$
(when $S=0$) and $\widetilde{R}_{k}\left(\mathbf{H},\widetilde{\mathbf{V}}\right)\geq\mu_{\pi_{k}}$
(when $S=1$) in (\ref{eq:Fset}) essentially guarantee that the playback
process is free from interruption.

Then the joint cache and power control problem is formulated as:
\begin{eqnarray}
\mathcal{P}:\:\underset{\mathbf{q}\in\mathcal{D}_{\mathbf{q}},\mathcal{V}}{\textrm{min}}\:\textrm{E}\left[\overline{P}_{\pi}\left(\mathbf{q},\mathcal{V}\right)\right]\;\;\;\;\;\;\;\;\;\;\;\;\;\;\;\;\;\;\;\;\;\;\;\;\;\;\;\;\;\;\;\;\;\;\;\;\;\;\;\;\;\;\nonumber \\
\textrm{s.t}.\: V\left(\pi,\mathbf{H}\right)\in\mathcal{D}_{\mathbf{v}}\left(\pi,\mathbf{H}\right),\widetilde{\mathbf{V}}\left(\pi,\mathbf{H}\right)\in\mathcal{D}_{\widetilde{\mathbf{v}}}\left(\pi,\mathbf{H}\right),\textrm{w.p.1},\label{eq:ratecon}
\end{eqnarray}
where the expectation is taken w.r.t. the distribution of $\pi$.
Note that in constraint (\ref{eq:ratecon}), the instantaneous rate
requirement is satisfied with probability one because it is impossible
to guarantee a fixed data rate for all realizations of $\mathbf{H}$. 

In general, problem $\mathcal{P}$ may not even be feasible (i.e.
$\textrm{E}\left[\overline{P}_{\pi}\left(\mathbf{q},\mathcal{V}\right)\right]$
is not bounded when constraint (\ref{eq:ratecon}) is satisfied).
However, the following proposition ensures that $\mathcal{P}$ is
feasible.
\begin{prop}
[Feasibility of $\mathcal{P}$]\label{prop:Feasibility-of-P}There
exists $\mathcal{V}=\left\{ \mathbf{V}\left(\pi,\mathbf{H}\right),\widetilde{\mathbf{V}}\left(\pi,\mathbf{H}\right):\forall\pi,\mathbf{H}\right\} $
such that $\mathbf{V}\left(\pi,\mathbf{H}\right)\in\mathcal{D}_{\mathbf{v}}\left(\pi,\mathbf{H}\right),\:\widetilde{\mathbf{V}}\left(\pi,\mathbf{H}\right)\in\mathcal{D}_{\widetilde{\mathbf{v}}}\left(\pi,\mathbf{H}\right)$
with probability one and $\textrm{E}\left[\overline{P}_{\pi}\left(\mathbf{q},\mathcal{V}\right)\right]$
is bounded.
\end{prop}

Please refer to Appendix \ref{sub:Proof-of-Proposition-PF} for the
proof.

\subsection{Problem Decomposition}

Problem $\mathcal{P}$ is a non-convex stochastic optimization problem.
We first decompose it into simpler subproblems. For convenience, define
\begin{equation}
\mathcal{H}_{F}\triangleq\left\{ \mathbf{H}:\:\textrm{Tr}\left(\mathbf{H}_{m,k,k}\mathbf{H}_{m,k,k}^{\dagger}\right)>0,\forall m,k\right\} .\label{eq:DefHF}
\end{equation}
According to the analysis in Appendix \ref{sub:Proof-of-Proposition-PF},
for any $\pi$ and $\mathbf{H}\in\mathcal{H}_{F}$, there exists $\mathbf{V},\widetilde{\mathbf{V}}$
such that $\mathbf{V}\in\mathcal{D}_{\mathbf{v}}\left(\pi,\mathbf{H}\right),\:\widetilde{\mathbf{V}}\in\mathcal{D}_{\widetilde{\mathbf{v}}}\left(\pi,\mathbf{H}\right)$
and $P\left(\mathbf{V}\right),\widetilde{P}\left(\widetilde{\mathbf{V}}\right)$
are bounded. Moreover, we have $\textrm{Pr}\left[\mathbf{H}\in\mathcal{H}_{F}\right]=1$.
Then by exploiting the timescale separations of the optimization variables,
problem $\mathcal{P}$ can be decomposed into the following families
of subproblems.

\textbf{Subproblem 1} (\textbf{Short-term Coordinated MIMO Precoding
}for\textbf{ }given $\pi$, $\mathbf{H}\in\mathcal{H}_{F}$ and $S=0$):
\[
\mathcal{P}_{S}\left(\pi,\mathbf{H}\right):\:\underset{\mathbf{V}}{\textrm{min}}\: P\left(\mathbf{V}\right),\textrm{s.t}.\:\mathbf{V}\in\mathcal{D}_{\mathbf{v}}\left(\pi,\mathbf{H}\right).
\]

\textbf{Subproblem 2} (\textbf{Short-term CoMP Precoding }for given
$\pi$, $\mathbf{H}\in\mathcal{H}_{F}$ and $S=1$):
\[
\widetilde{\mathcal{P}}_{S}\left(\pi,\mathbf{H}\right):\:\underset{\widetilde{\mathbf{V}}}{\textrm{min}}\:\widetilde{P}\left(\widetilde{\mathbf{V}}\right),\textrm{s.t}.\:\widetilde{\mathbf{V}}\in\mathcal{D}_{\widetilde{\mathbf{v}}}\left(\pi,\mathbf{H}\right).
\]

\textbf{Subproblem 3 (Long-term Cache Control }for given\textbf{ $\mathcal{V}$)}:
\begin{equation}
\mathcal{P}_{L}\left(\mathcal{V}\right)\::\underset{\mathbf{q}\in\mathcal{D}_{\mathbf{q}}}{\textrm{min}}\:\psi\left(\mathbf{q},\mathcal{V}\right)\triangleq\textrm{E}\left[\overline{P}_{\pi}\left(\mathbf{q},\mathcal{V}\right)\right].\label{eq:KKTq}
\end{equation}

The relationship between problem $\mathcal{P}$ and the subproblems
is summarized as follows. For given $\pi,\mathbf{H}\in\mathcal{H}_{F}$,
let $\mathbf{V}^{\star}\left(\pi,\mathbf{H}\right)$ be the optimal
solution of $\mathcal{P}_{S}\left(\pi,\mathbf{H}\right)$ and $\widetilde{\mathbf{V}}^{\star}\left(\pi,\mathbf{H}\right)$
be the optimal solution of $\widetilde{\mathcal{P}}_{S}\left(\pi,\mathbf{H}\right)$.
For given $\pi,\mathbf{H}\notin\mathcal{H}_{F}$, let $\mathbf{V}^{\star}\left(\pi,\mathbf{H}\right)=\left\{ \mathbf{V}_{m,k}^{\star}\left(\pi,\mathbf{H}\right)=\mathbf{0}:\forall m,k\right\} $
and $\widetilde{\mathbf{V}}^{\star}\left(\pi,\mathbf{H}\right)=\left\{ \widetilde{\mathbf{V}}_{m,k}^{\star}\left(\pi,\mathbf{H}\right)=\mathbf{0}:\forall m,k\right\} $.
Let $\mathbf{q}^{\star}$ be the optimal solution of $\mathcal{P}_{L}\left(\mathcal{V}^{\star}\right)$,
where $\mathcal{V}^{\star}=\left\{ \mathbf{V}^{\star}\left(\pi,\mathbf{H}\right),\widetilde{\mathbf{V}}^{\star}\left(\pi,\mathbf{H}\right):\forall\pi,\mathbf{H}\right\} $.
Then $\left(\mathbf{q}^{\star},\mathcal{V}^{\star}\right)$ is the
optimal solution of $\mathcal{P}$.%

The above three subproblems are still non-trivial. Although the gradient
projection (GP) method \cite{Bertsekas_book99_NProgramming} is usually
used to find a stationary point for a constrained non-convex problem,
it cannot be applied to solve $\mathcal{P}_{S}\left(\pi,\mathbf{H}\right)$
because $\mathcal{D}_{\mathbf{v}}\left(\pi,\mathbf{H}\right)$ is
a non-convex set and calculating the projection of $\mathbf{V}$ on
the feasible set $\mathcal{D}_{\mathbf{v}}\left(\pi,\mathbf{H}\right)$
is also a non-convex problem. Similar observations can also be made
for $\widetilde{\mathcal{P}}_{S}\left(\pi,\mathbf{H}\right)$. In
Section \ref{sec:Low-Complexity-Power}, we generalize the WMMSE approach
in \cite{Luo_TSP11_WMMSE} to obtain a polynomial complexity algorithm
which converges to a stationary point of the short-term precoding
problem $\mathcal{P}_{S}\left(\pi,\mathbf{H}\right)$ or $\widetilde{\mathcal{P}}_{S}\left(\pi,\mathbf{H}\right)$.
In Section \ref{sec:Asymptotically-Optimal-out_Solution}, we exploit
the hidden-convexity and propose a robust stochastic subgradient algorithm
to solve $\mathcal{P}_{L}\left(\mathcal{V}\right)$.

\section{Short-term Precoding Solutions for $\mathcal{P}_{S}\left(\pi,\mathbf{H}\right)$
and $\widetilde{\mathcal{P}}_{S}\left(\pi,\mathbf{H}\right)$\label{sec:Low-Complexity-Power}}

Problem $\mathcal{P}_{S}\left(\pi,\mathbf{H}\right)$ is a sum power
minimization problem under individual rate constraints in parallel
interference networks. In \cite{Luo_TSP11_WMMSE}, a WMMSE algorithm
was proposed to find a stationary point for the weighted sum-rate
maximization problem (WSRMP) in MIMO interfering broadcast channels
under per-BS power constraints. In the following, the WMMSE algorithm
is generalized to solve $\mathcal{P}_{S}\left(\pi,\mathbf{H}\right)$
and $\widetilde{\mathcal{P}}_{S}\left(\pi,\mathbf{H}\right)$.

\subsection{An Equivalent Problem under Weighted MSE Constraint}

Consider a sum power minimization problem under individual weighted
MSE constraints:{\small{
\begin{eqnarray}
\underset{\left\{ \mathbf{W},\mathbf{U},\mathbf{V}\right\} }{\textrm{min}}\: P\left(\mathbf{V}\right)\triangleq\sum_{k=1}^{K}\sum_{m=1}^{M}\textrm{Tr}\left(\mathbf{V}_{m,k}\mathbf{V}_{m,k}^{\dagger}\right)\;\;\;\;\;\;\;\label{eq:PMSE}\\
\textrm{s.t.}\:\frac{\sum_{m=1}^{M}\left(\textrm{Tr}\left(\mathbf{\mathbf{W}}_{m,k}\mathbf{E}_{m,k}\right)-\textrm{log}\left|\mathbf{\mathbf{W}}_{m,k}\right|\right)}{M}\leq d-\overline{\mu}_{\pi_{k}},\forall k,\nonumber 
\end{eqnarray}
}}where $\mathbf{W}=\left\{ \mathbf{\mathbf{W}}_{m,k}\succeq\mathbf{0}:\forall m,k\right\} $
is a set of weight matrices; $\mathbf{U}=\left\{ \mathbf{U}_{m,k}:\forall m,k\right\} $
is the set of all receiving matrices; $\overline{\mu}_{\pi_{k}}=\frac{\mu_{\pi_{k}}\ln2}{B_{W}}$
and {\small{
\begin{eqnarray}
\mathbf{E}_{m,k}=\left(\mathbf{I}-\mathbf{U}_{m,k}^{\dagger}\mathbf{H}_{m,k,k}\mathbf{V}_{m,k}\right)\left(\mathbf{I}-\mathbf{U}_{m,k}^{\dagger}\mathbf{H}_{m,k,k}\mathbf{V}_{m,k}\right)^{\dagger}\nonumber \\
+\mathbf{U}_{m,k}^{\dagger}\mathbf{\Omega}_{m,k}\mathbf{U}_{m,k},\;\;\;\;\;\;\;\;\;\;\;\;\;\;\;\;\;\;\;\;\;\;\;\;\;\;\;\;\;\;\;\;\;\;\;\;\;\;\;\;\;\;\;\;\;\;\;\;\;\label{eq:Emk}
\end{eqnarray}
}}is the MSE matrix of user $k$ on subcarrier $m$. The following
theorem establishes the equivalence between $\mathcal{P}_{S}\left(\pi,\mathbf{H}\right)$
and Problem (\ref{eq:PMSE}).
\begin{thm}
[Equivalence between $\mathcal{P}_{S}\left(\pi,\mathbf{H}\right)$
and (\ref{eq:PMSE})]\label{thm:Equivalence-between-PSandPMSE}For
given $\pi,\mathbf{H}\in\mathcal{H}_{F}$, let $\left(\mathbf{W}^{\star},\mathbf{U}^{\star},\mathbf{V}^{\star}\right)$
denote the optimal solution of Problem (\ref{eq:PMSE}). Then $\mathbf{V}^{\star}$
is also the optimal solution of $\mathcal{P}_{S}\left(\pi,\mathbf{H}\right)$.
\end{thm}

Please refer to Appendix \ref{sub:Proof-of-TheoremEquiPS11} for the
proof.

Hence we only need to solve Problem (\ref{eq:PMSE}), which is convex
in each of the optimization variables $\mathbf{W},\mathbf{U},\mathbf{V}$.
We can use the block coordinate decent method to solve (\ref{eq:PMSE}).
First, for fixed $\mathbf{V},\mathbf{U}$, the optimal $\mathbf{W}$
is given by $\mathbf{W}_{m,k}=\mathbf{E}_{m,k}^{-1},\forall m,k$.
Second, for fixed $\mathbf{V},\mathbf{W}$, the optimal $\mathbf{U}$
is given by the MMSE receiver in (\ref{eq:MMSEu}). Finally, for fixed
$\mathbf{U},\mathbf{W}$, Problem (\ref{eq:PMSE}) is a convex quadratic
optimization problem which can be solved using the Lagrange dual method
as will be elaborated in the next subsection.

\subsection{Lagrange dual method for solving Problem (\ref{eq:PMSE}) with fixed
$\mathbf{U},\mathbf{W}$}

The Lagrange function of Problem (\ref{eq:PMSE}) with fixed $\mathbf{U},\mathbf{W}$
is given by
\begin{eqnarray*}
L\left(\boldsymbol{\lambda},\mathbf{V},\mathbf{U},\mathbf{W}\right)=\sum_{k=1}^{K}\sum_{m=1}^{M}\textrm{Tr}\left(\mathbf{V}_{m,k}\mathbf{V}_{m,k}^{\dagger}\right)+\;\;\;\;\;\;\;\;\;\;\;\;\;\;\;\;\;\;\;\\
\sum_{k=1}^{K}\lambda_{k}\left(\frac{\sum_{m=1}^{M}\left(\textrm{Tr}\left(\mathbf{\mathbf{W}}_{m,k}\mathbf{E}_{m,k}\right)-\textrm{log}\left|\mathbf{\mathbf{W}}_{m,k}\right|\right)}{M}-d+\overline{\mu}_{\pi_{k}}\right),
\end{eqnarray*}
where $\boldsymbol{\lambda}=\left[\lambda_{k}\right]_{k=1,...,K}\in\mathbb{R}_{+}^{K}$
is the Lagrange multiplier vector. The dual function of Problem (\ref{eq:PMSE})
with fixed $\mathbf{U},\mathbf{W}$ is
\begin{equation}
J\left(\boldsymbol{\lambda}\right)=\underset{\mathbf{V}}{\textrm{min}}\: L\left(\boldsymbol{\lambda},\mathbf{V},\mathbf{U},\mathbf{W}\right).\label{eq:DualFun}
\end{equation}
The minimization problem in (\ref{eq:DualFun}) can be decomposed
into $M$ independent problems as 
\begin{equation}
\underset{\left\{ \mathbf{V}_{m,k}\right\} }{\textrm{min}}\:\sum_{k=1}^{K}\textrm{Tr}\left(\mathbf{V}_{m,k}\mathbf{V}_{m,k}^{\dagger}\right)+\sum_{k=1}^{K}\frac{\lambda_{k}}{M}\textrm{Tr}\left(\mathbf{\mathbf{W}}_{m,k}\mathbf{E}_{m,k}\right),\:\forall m.\label{eq:maxLm}
\end{equation}
For fixed $\boldsymbol{\lambda}$, Problem (\ref{eq:maxLm}) has a
closed-form solution given by 
\begin{eqnarray}
 &  & \mathbf{V}_{m,k}^{*}\left(\boldsymbol{\lambda}\right)\nonumber \\
 & = & \left(\sum_{n=1}^{K}\frac{\lambda_{n}}{M}\mathbf{H}_{m,n,k}^{\dagger}\mathbf{U}_{m,n}\mathbf{\mathbf{W}}_{m,n}\mathbf{U}_{m,n}^{\dagger}\mathbf{H}_{m,n,k}+\mathbf{I}\right)^{-1}\nonumber \\
 &  & \times\frac{\lambda_{k}}{M}\mathbf{H}_{m,k,k}^{\dagger}\mathbf{U}_{m,k}\mathbf{\mathbf{W}}_{m,k},\:\forall k.\label{eq:optqk}
\end{eqnarray}
Since Problem (\ref{eq:PMSE}) with fixed $\mathbf{U},\mathbf{W}$
is a convex quadratic optimization problem, the optimal solution is
given by $\mathbf{V}^{*}\left(\boldsymbol{\lambda}^{*}\right)=\left\{ \mathbf{V}_{m,k}^{*}\left(\boldsymbol{\lambda}^{*}\right):\forall m,k\right\} $,
where $\boldsymbol{\lambda}^{*}$ is the optimal solution of the dual
problem
\begin{equation}
\underset{\boldsymbol{\lambda}}{\textrm{max}}\: J\left(\boldsymbol{\lambda}\right),\:\textrm{s.t.}\:\boldsymbol{\lambda}\geq\mathbf{0}.\label{eq:mingfun}
\end{equation}
The dual function $J\left(\boldsymbol{\lambda}\right)$ is concave
and it can be verified that{\small{
\begin{equation}
\left[\frac{\sum_{m=1}^{M}\left(\textrm{Tr}\left(\mathbf{\mathbf{W}}_{m,k}\mathbf{E}_{m,k}^{*}\left(\boldsymbol{\lambda}\right)\right)-\textrm{log}\left|\mathbf{\mathbf{W}}_{m,k}\right|\right)}{M}-d+\overline{\mu}_{\pi_{k}}\right]_{k=1,...,K}\label{eq:subJ}
\end{equation}
}}is a subgradient of $J\left(\boldsymbol{\lambda}\right)$, where
$\mathbf{E}_{m,k}^{*}\left(\boldsymbol{\lambda}\right)$ is obtained
from (\ref{eq:Emk}) with $\mathbf{V}_{m,k}=\mathbf{V}_{m,k}^{*}\left(\boldsymbol{\lambda}\right)$.
Hence, the standard subgradient based methods such as the subgradient
algorithm in \cite{Boyd_03note_Subgradient} or the Ellipsoid method
in \cite{Boyd_04Book_Convex_optimization} can be used to solve the
optimal solution $\boldsymbol{\lambda}^{*}$ of the dual problem in
(\ref{eq:mingfun}).

\subsection{Overall Algorithm for Solving $\mathcal{P}_{S}\left(\pi,\mathbf{H}\right)$
and $\widetilde{\mathcal{P}}_{S}\left(\pi,\mathbf{H}\right)$}

The overall algorithm (named Algorithm SP) for solving $\mathcal{P}_{S}\left(\pi,\mathbf{H}\right)$
is summarized in Table \ref{tab:AlgPS}. Note that using the Matrix
Inversion Lemma, it can be shown that $\mathbf{E}_{m,k}^{-1}=\left(\mathbf{I}-\mathbf{U}_{m,k}^{\dagger}\mathbf{H}_{m,k,k}\mathbf{V}_{m,k}\right)^{-1}$
if $\mathbf{U}_{m,k}$ is the MMSE receiver given in the step 1 of
Algorithm SP. Hence in step 2, we let $\mathbf{\mathbf{W}}_{m,k}=\left(\mathbf{I}-\mathbf{U}_{m,k}^{\dagger}\mathbf{H}_{m,k,k}\mathbf{V}_{m,k}\right)^{-1}$.
The following theorem shows that Algorithm SP converges to a stationary
point of $\mathcal{P}_{S}\left(\pi,\mathbf{H}\right)$.
\begin{thm}
[Convergence of Alg. SP]\label{thm:Convergence-of-AlgPS}For given
$\pi,\mathbf{H}\in\mathcal{H}_{F}$, any limit point $\left(\mathbf{W}^{*},\mathbf{U}^{*},\mathbf{V}^{*}\right)$
of the iterates generated by Algorithm SP is a stationary point of
Problem (\ref{eq:PMSE}), and the corresponding $\mathbf{V}^{*}$
is a stationary point of $\mathcal{P}_{S}\left(\pi,\mathbf{H}\right)$.
\end{thm}

Please refer to Appendix \ref{sub:Proof-of-TheoremconvPS} for the
proof.

Problem $\widetilde{\mathcal{P}}_{S}\left(\pi,\mathbf{H}\right)$
is a sum power minimization problem under individual rate constraints
in parallel broadcast channel, which can be viewed as a parallel interference
network with the cross link channel equal to the direct link channel.
Hence, Algorithm SP can also be used to find a stationary point $\widetilde{\mathbf{V}}^{*}\left(\pi,\mathbf{H}\right)$
of $\widetilde{\mathcal{P}}_{S}\left(\pi,\mathbf{H}\right)$ by replacing
$\mathbf{H}_{m,k,n},\forall n$, $\mathbf{V}_{m,k}$, $\mathbf{U}_{m,k}$,
$\mathbf{\Omega}_{m,k}$ respectively with $\widetilde{\mathbf{H}}_{m,k}$,
$\widetilde{\mathbf{V}}_{m,k}$, $\widetilde{\mathbf{U}}_{m,k}$,
$\widetilde{\mathbf{\Omega}}_{m,k}$ and using the following initial
point:
\begin{equation}
\left[\widetilde{\mathbf{V}}_{m,k}\right]_{i,j}=\begin{cases}
\left[\mathbf{V}_{m,k}^{*}\left(\pi,\mathbf{H}\right)\right]_{i-\left(k-1\right)N_{T},j}, & i\in\mathcal{I}_{k},j\in\left[1,d\right]\\
0, & \textrm{otherwise}
\end{cases}\label{eq:initVta}
\end{equation}
for all $m,k$, where $\mathcal{I}_{k}=\left[\left(k-1\right)N_{T}+1,kN_{T}\right]$;
and $\mathbf{V}^{*}\left(\pi,\mathbf{H}\right)=\left\{ \mathbf{V}_{m,k}^{*}\left(\pi,\mathbf{H}\right),\forall m,k\right\} $
is the stationary point of $\mathcal{P}_{S}\left(\pi,\mathbf{H}\right)$
found by Algorithm SP. The initial point in (\ref{eq:initVta}) is
chosen to ensure that $\widetilde{P}\left(\widetilde{\mathbf{V}}^{*}\left(\pi,\mathbf{H}\right)\right)\leq P\left(\mathbf{V}^{*}\left(\pi,\mathbf{H}\right)\right)$,
which is the key to prove the convexity of the long-term cache control
problem $\mathcal{P}_{L}\left(\mathcal{V}^{*}\right)$ in Lemma \ref{lem:Convexity-of-PL}.

\begin{table}
\caption{\label{tab:AlgPS}Algorithm SP (for finding a stationary point of
$\mathcal{P}_{S}\left(\pi,\mathbf{H}\right))$}

\centering{}%
\begin{tabular}{l}
\hline 
\textbf{\small{Initialize}}{\small{ $\mathbf{V}_{m,k}$'s such that
$R_{k}\left(\mathbf{H},\mathbf{V}\right)\geq\overline{\mu}_{\pi_{k}},\:\forall k$.}}\tabularnewline
\textbf{\small{Step 1:}}{\small{ Let $\mathbf{U}_{m,k}=\left(\mathbf{\Omega}_{m,k}+\mathbf{H}_{m,k,k}\mathbf{V}_{m,k}\mathbf{V}_{m,k}^{\dagger}\mathbf{H}_{m,k,k}^{\dagger}\right)^{-1}$}}\tabularnewline
{\small{$\;\;\;\;\;\;\;\;\;\;\;\;\;\;\;\;\;\;\;\;\;\;\;\;\;\;\;\;\;\;\;\;\;\times\mathbf{H}_{m,k,k}\mathbf{V}_{m,k},\forall m,k.$}}\tabularnewline
\textbf{\small{Step 2:}}{\small{ Let $\mathbf{\mathbf{W}}_{m,k}=\left(\mathbf{I}-\mathbf{U}_{m,k}^{\dagger}\mathbf{H}_{m,k,k}\mathbf{V}_{m,k}\right)^{-1},\forall m,k.$}}\tabularnewline
\textbf{\small{Step 3:}}{\small{ Let $\mathbf{V}_{m,k}=\mathbf{V}_{m,k}^{*}\left(\boldsymbol{\lambda^{*}}\right),\forall m,k$,
where $\boldsymbol{\lambda^{*}}$ is the}}\tabularnewline
{\small{$\;\;\;\;\;\;\;\;\;\;\;\;$optimal solution of (\ref{eq:mingfun})
which can be solved using, e.g.,}}\tabularnewline
{\small{$\;\;\;\;\;\;\;\;\;\;\;\;$the subgradient algorithm in \cite{Boyd_03note_Subgradient}
or the Ellipsoid }}\tabularnewline
{\small{$\;\;\;\;\;\;\;\;\;\;\;\;$method in \cite{Boyd_04Book_Convex_optimization}
with the subgradient of $J\left(\boldsymbol{\lambda}\right)$ given
in }}\tabularnewline
{\small{$\;\;\;\;\;\;\;\;\;\;\;\;$(\ref{eq:subJ}) and $\mathbf{V}_{m,k}^{*}\left(\boldsymbol{\lambda}\right)$
is given in (\ref{eq:optqk}).}}\tabularnewline
\textbf{\small{Return to Step 1 until convergence.}}\tabularnewline
\hline 
\end{tabular}
\end{table}

\section{Long term Cache Control for $\mathcal{P}_{L}\left(\mathcal{V}^{*}\right)$\label{sec:Asymptotically-Optimal-out_Solution}}

The following lemma shows that $\mathcal{P}_{L}\left(\mathcal{V}^{*}\right)$
is a convex stochastic optimization problem as long as $\mathcal{V}^{*}$
is a stationary point of $\mathcal{P}_{S}\left(\pi,\mathbf{H}\right)$
and $\widetilde{\mathcal{P}}_{S}\left(\pi,\mathbf{H}\right)$ found
by Algorithm SP. 
\begin{lem}
[Hidden Convexity of $\mathcal{P}_{L}\left(\mathcal{V}^{*}\right)$]\label{lem:Convexity-of-PL}For
any $\pi$ and $\mathbf{H}\in\mathcal{H}_{F}$, let $\mathbf{V}^{*}\left(\pi,\mathbf{H}\right)$
be the stationary point of $\mathcal{P}_{S}\left(\pi,\mathbf{H}\right)$
found by Algorithm SP and let $\widetilde{\mathbf{V}}^{*}\left(\pi,\mathbf{H}\right)$
be the stationary point of $\widetilde{\mathcal{P}}_{S}\left(\pi,\mathbf{H}\right)$
found by Algorithm SP with the initial point given in (\ref{eq:initVta}).
For given $\pi,\mathbf{H}\notin\mathcal{H}_{F}$, let $\mathbf{V}^{*}\left(\pi,\mathbf{H}\right)=\left\{ \mathbf{V}_{m,k}^{*}\left(\pi,\mathbf{H}\right)=\mathbf{0}:\forall m,k\right\} $
and $\widetilde{\mathbf{V}}^{*}\left(\pi,\mathbf{H}\right)=\left\{ \widetilde{\mathbf{V}}_{m,k}\left(\pi,\mathbf{H}\right)=\mathbf{0}:\forall m,k\right\} $.
Then $\mathcal{P}_{L}\left(\mathcal{V}^{*}\right)$ with $\mathcal{V}^{*}=\left\{ \mathbf{V}^{*}\left(\pi,\mathbf{H}\right),\widetilde{\mathbf{V}}^{*}\left(\pi,\mathbf{H}\right):\forall\pi,\mathbf{H}\right\} $
is a convex stochastic optimization problem.
\end{lem}

Please refer to Appendix \ref{sub:Proof-of-PropositionCPL} for the
proof. 

Hence, we propose a stochastic subgradient algorithm 
which is able to converge to the optimal solution of $\mathcal{P}_{L}\left(\mathcal{V}^{*}\right)$
without knowing the distribution of $\pi$. 

For notation convenience, let $T_{i}$ denote the time slot when $\pi$
changes for the $i$-th time, i.e., $\pi$ remains constant for each
time interval $\left[T_{i},T_{i+1}-1\right]$ and changes at the boundary
of each time interval. Then the following lemma gives a noisy unbiased
subgradient of the objective function
$\psi\left(\mathbf{q},\mathcal{V}^{*}\right)$.
\begin{lem}
[Noisy unbiased subgradient of $\psi\left(\mathbf{q},\mathcal{V}^{*}\right)$]\label{lem:subconvcond}Let
$\pi^{(i)}$ denote the URP for the $i$-th time interval $\left[T_{i},T_{i+1}-1\right]$.
At time slot $T_{i+1}-1$, a noisy unbiased subgradient of $\psi\left(\mathbf{q},\mathcal{V}^{*}\right)$
at $\mathbf{q}\in\mathcal{D}_{\mathbf{q}}$, denoted by $\hat{\nabla}\psi^{(i)}\left(\mathbf{q}\right)=\left[\widehat{\frac{\partial\psi^{(i)}}{\partial q_{1}}},...,\widehat{\frac{\partial\psi^{(i)}}{\partial q_{L}}}\right]^{T}$,
is given by
\begin{eqnarray}
\widehat{\frac{\partial\psi^{(i)}}{\partial q_{l}}} & = & 1\left(l=\pi_{k^{*}}^{(i)}\right)\left(\frac{\sum_{t\in\widetilde{\mathcal{T}}^{(i)}}\widetilde{P}\left(\widetilde{\mathbf{V}}^{*}\left(\pi^{(i)},\mathbf{H}\left(t\right)\right)\right)}{\left|\widetilde{\mathcal{T}}^{(i)}\right|}\right.\nonumber \\
 &  & \left.-\frac{\sum_{t\in\mathcal{T}^{(i)}}P\left(\mathbf{V}^{*}\left(\pi^{(i)},\mathbf{H}\left(t\right)\right)\right)}{\left|\mathcal{T}^{(i)}\right|}\right),\forall l\label{eq:subfai}
\end{eqnarray}
where $k^{*}$ is any index satisfying $q_{\pi_{k^{*}}}=\underset{1\leq k\leq K}{\textrm{min}}\left\{ q_{\pi_{k}}\right\} $;
$\widetilde{\mathcal{T}}^{(i)},\mathcal{T}^{(i)}$ are any two non-empty
sets of time slot index in $\left[T_{i},T_{i+1}-1\right]$.
\end{lem}

Please refer to Appendix \ref{sub:Proof-of-Lemmasubconv} for the
proof. In practical implementation, we can choose $\mathcal{T}^{(i)}=\left\{ t\in\left[T_{i},T_{i+1}-1\right]:\: S\left(t\right)=0\right\} $
and $\widetilde{\mathcal{T}}^{(i)}=\left\{ t\in\left[T_{i},T_{i+1}-1\right]:\: S\left(t\right)=1\right\} $
to avoid the extra computation for the subgradient calculation, providing
that the two sets are not empty.

After obtaining the noisy unbiased subgradient using (\ref{eq:subfai}),
$\mathbf{q}$ is updated using the following subgradient projection
method
\begin{equation}
\mathbf{q}^{(i+1)}=\underset{\mathbf{q}\in\mathcal{D}_{\mathbf{q}}}{\textrm{argmin}}\left\Vert \mathbf{q}^{(i)}-\sigma^{(i)}\hat{\nabla}\psi^{(i)}\left(\mathbf{q}^{(i)}\right)-\mathbf{q}\right\Vert ^{2},\label{eq:Projq}
\end{equation}
where $\sigma^{(i)}>0$ is the step size for the $i$-th update. The
projection problem in (\ref{eq:Projq}) is a convex quadratic optimization
problem and can be easily solved by the standard convex optimization
methods \cite{Boyd_04Book_Convex_optimization}.

Finally, the overall algorithm is summarized in Table \ref{tab:Algq}
and the global convergence is established in the following Theorem.
\begin{table}
\caption{\label{tab:Algq}Algorithm LC (for Solving Problem $\mathcal{P}_{L}\left(\mathcal{V}^{*}\right)$)}

\centering{}%
\begin{tabular}{l}
\hline 
\textbf{\small{Initialization: }}{\small{Let $\mathbf{q}^{(0)}=\mathbf{0}$
and $i=0$.}}\tabularnewline
\textbf{\small{Step 1:}}{\small{ At time slot $t=T_{i+1}-1$, calculate
a noisy}}\tabularnewline
{\small{$\;\;\;\;\;\;\;\;$unbiased subgradient of $\psi\left(\mathbf{q}^{(i)},\mathcal{V}^{*}\right)$
using (\ref{eq:subfai}).}}\tabularnewline
\textbf{\small{Step 2:}}{\small{ Choose proper step size $\sigma^{(i)}>0$
and }}\tabularnewline
{\small{$\;\;\;\;\;\;\;\;$obtain $\mathbf{q}^{(i+1)}$ using (\ref{eq:Projq}).}}\tabularnewline
\textbf{\small{Step 3:}}{\small{ Let $i=i+1$ and return to Step 1.}}\tabularnewline
\hline 
\end{tabular}
\end{table}

\begin{thm}
[Convergence of Algorithm LC]If the step sizes $\sigma^{(i)}>0$
in Algorithm LC satisfies: 1) $\sum_{i=1}^{\infty}\left(\sigma^{(i)}\right)^{2}<\infty$;
2) $\sum_{i=1}^{\infty}\sigma^{(i)}=\infty$, then Algorithm LC converges
to an optimal solution $\mathbf{q}^{*}$ of Problem $\mathcal{P}_{L}\left(\mathcal{V}^{*}\right)$
with probability 1.
\end{thm}

The convergence proof follows directly from Lemma \ref{lem:Convexity-of-PL},
Lemma \ref{lem:subconvcond} and \cite[Theorem 3.3]{Ram_SIAM2009_SubgradConv}. 
\begin{rem}
\label{rem:Hiid}The proof of \cite[Theorem 3.3]{Ram_SIAM2009_SubgradConv}
requires \cite[Assumption 2]{Ram_SIAM2009_SubgradConv}, which holds
when $\mathbf{H}$ is i.i.d. across the time slot index $t$.
\end{rem}

\begin{rem}
Algorithm LC only requires observation of the realizations of URP
$\pi$ in every update. However, it does not need to know the \textbf{statistics}
\textbf{of the URP}.%
{} 
\end{rem}

\section{Implementation Considerations}

\subsection{Summary of the Overall Solution}

\begin{figure}
\begin{centering}
\includegraphics[width=85mm]{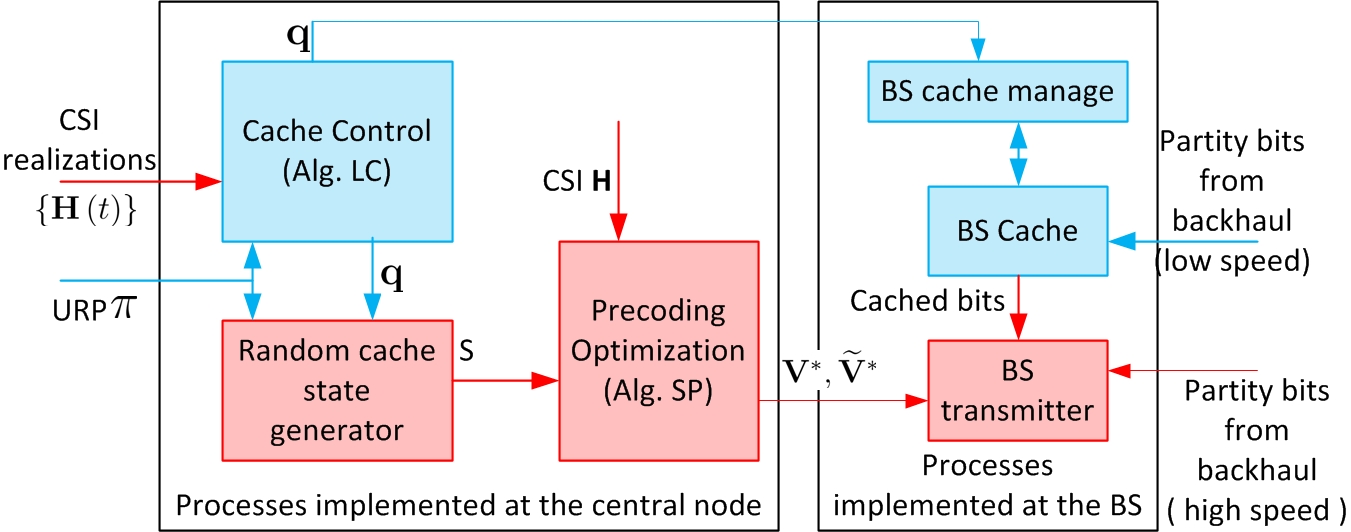}
\par\end{centering}

\caption{\label{fig:moduleconn}{\small{Summary of overall solution and the
inter-relationship of the algorithm components. The blue / red blocks
represent long timescale / short timescale processes. The blue / red
arrows represent long-term / short-term signaling.}}}
\end{figure}

Fig. \ref{fig:moduleconn} summarizes the overall solution and the
inter-relationship of the algorithm components. The solutions are
divided into \textit{long timescale process} and \textit{short timescale
process}. The long timescale processing consists of Algorithm LC (cache
control) and BS cache management. The short timescale processing consists
of random cache state generator and Algorithm SP (precoding optimization).
The precoding optimization and cache control processes are implemented
at the central node, while the cache management process is implemented
at each BS. Whenever the URP $\pi$ changes, the updated cache control
vector $\mathbf{q}$ is computed from the central node and passed
to each BS. Then the BS cache management updates the BS cache according
to $\mathbf{q}$. Specifically, if the cached parity bits for each
segment of the $l$-th media file is less than $q_{l}L_{S}$ bits,
it will request new parity bits from the backhaul. Otherwise, it will
drop some cached parity bits. At each time slot, the cache state $S$
is generated from the random cache state generator using $\mathbf{q}$
and $\pi$, and the CSI $\mathbf{H}$ is obtained via feedback from
the users. Furthermore, at each time slot, the precoding matrices
$\mathbf{V},\widetilde{\mathbf{V}}$ are determined at the central
node based on the CSI $\mathbf{H}$ and cache state $S$. Then they
are sent to the BSs for MIMO transmission.

\subsection{Complexity in Computation and Signaling}

\subsubsection*{Computation Complexity }

The complexity of the precoding optimization Algorithm SP is similar
to the WMMSE algorithm in \cite{Luo_TSP11_WMMSE} and is polynomial
w.r.t. the number of users and antennas at each node. Please refer
to \cite{Luo_TSP11_WMMSE} for details. The long term cache control
only needs to do a simple subgradient projection update in (\ref{eq:Projq})
for each realization of $\pi$ and thus the complexity is extremely
low.

\subsubsection*{Control Signaling Overhead}

The short term control signaling overhead is similar to the conventional
coordinated MIMO schemes and can be supported by the modern wireless
systems such as LTE \cite{LTE}. At each time slot, user $k$ feedbacks
its direct and cross link CSI $\mathbf{H}_{m,k,n},\forall m,n$ to
the central node. Then the central node broadcasts the cache state
$S$ and precoding matrices $\mathbf{V}$ or $\widetilde{\mathbf{V}}$
to the BSs. The long term cache control signaling between the BSs
and central node is very small since $\mathbf{q}$ is only sent to
the BS once for each realization of $\pi$.

\subsubsection*{Average Backhaul Consumption}

For simplicity, we assume $\mu_{l}=\mu_{0},\forall l$. Suppose Algorithm
LC converges to $\mathbf{q}$. We first analyze the backhaul consumption
due to online media streaming. For each segment of the media file
requested by user $k$, there are only $\left(1-\textrm{min}_{k}\left\{ q_{\pi_{k}}\right\} \right)\mu_{0}T_{S}$
parity bits from the backhaul when the URP is $\pi$. Hence, the total
average backhaul consumption due to online media streaming for $K$
users is $\textrm{E}\left[K\left(1-\textrm{min}_{k}\left\{ q_{\pi_{k}}\right\} \right)\mu_{0}\right]$.
For the BS cache update process, each BS needs to obtain a total number
of $\sum_{l=1}^{L}q_{l}F_{l}$ parity bits from the backhaul. However,
these cache updates can be done offline with much smaller average
backhaul consumption compared with the backhaul consumption due to
online streaming because the popularity of media files changes very
slowly (e.g. new movies are usually posted on a weekly or monthly
timescale). Let $T_{C}$ denote the interval of the cache update process,
then the average backhaul consumption due to the offline BS cache
update is $K\sum_{l=1}^{L}q_{l}F_{l}/T_{C}$. Finally, the overall
average backhaul consumption (bps) is given by
\begin{equation}
R_{B}=\textrm{E}\left[K\left(1-\textrm{min}_{k}\left\{ q_{\pi_{k}}\right\} \right)\mu_{0}\right]+K\sum_{l=1}^{L}q_{l}F_{l}/T_{C}.\label{eq:OnlineBC}
\end{equation}
Table. \ref{tab:cacheupdateload} compares the average backhaul consumption
of different schemes. Impressively, the average backhaul consumption
is much smaller than the coordinated MIMO for only moderately large
BS cache size $B_{C}$.
\begin{table}
\begin{centering}
{\small{}}%
\begin{tabular}{|l|l|l|}
\hline 
 & {\small{Backhaul }} & {\small{Transmit power}}\tabularnewline
 & {\small{Consumption}} & {\small{performance}}\tabularnewline
\hline 
{\small{Proposed, $B_{C}=1.8$G}} & {\small{1.5868 Mbps}} & {\small{12.0937 dB}}\tabularnewline
{\small{Proposed, $B_{C}=1.2$G}} & {\small{5.7591 Mbps}} & {\small{13.4056 dB}}\tabularnewline
{\small{Coordinated MIMO}} & {\small{14 Mbps}} & {\small{15.2096 dB}}\tabularnewline
{\small{Conventional CoMP}} & {\small{98 Mbps}} & {\small{11.4782 dB}}\tabularnewline
{\small{Baseline 3, $B_{C}=1.2$G}} & {\small{9.3472 Mbps}} & {\small{14.2828 dB}}\tabularnewline
\hline 
\end{tabular}
\par\end{centering}{\small \par}

\caption{\label{tab:cacheupdateload}{\small{Backhaul consumption comparison
of different schemes. The corresponding average transmit power required
for QoS guarantee is also illustrated. We assume that the popularity
of the media files (i.e., the distribution of $\pi$) changes every
week, i.e., $T_{C}=1$ week. The system setup is given in Section
\ref{sec:Simulation-Results} and the results are evaluated under
edge user placement with a streaming rate of $\mu_{l}=\mu_{0}=2$Mbits/s,
$\forall l$. The backhaul consumption under normal user placement
is similar and is omitted for conciseness.}}}
\end{table}

\section{Simulation Results\label{sec:Simulation-Results}}

\begin{figure}
\begin{centering}
\includegraphics[width=55mm]{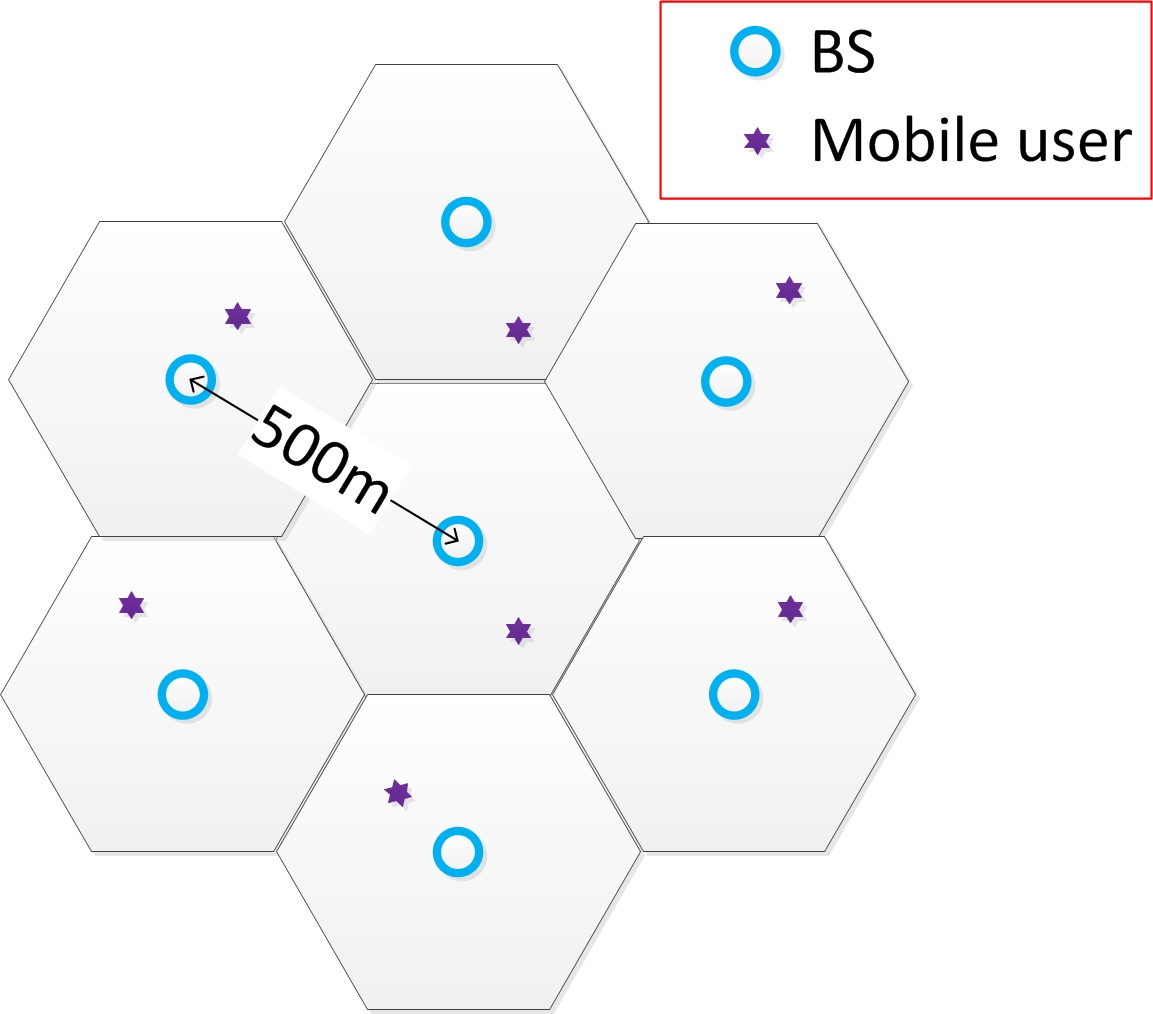}
\par\end{centering}

\caption{\label{fig:HetNet-topo}{\small{Topology of a multi-cell downlink
with $7$ cells.}}}
\end{figure}

Consider a media streaming system with $L=6$ media files and $K=7$
BS-user pairs. The BSs are arranged as in Fig. \ref{fig:HetNet-topo}
with inter-site distance of $500$m. We consider two different user
placements. In the \textit{normal user placement}, each user is uniformly
distributed within its cell under the restriction that the distance
between the user and the serving BS must be larger than%
\footnote{If the distance between a user and its serving BS is less than $80$m,
even the path gain of the strongest cross link is about 28dB smaller
than the path gain of the direct link. Such users are essentially
free from the inter-cell interference and we have excluded this uninteresting
degenerate case in the simulations.%
} $80$m. In the \textit{edge user placement}, each user is uniformly
distributed within its cell under the restriction that the distance
between the user and the serving BS must be larger than $180$m. Each
BS has $N_{T}=4$ antennas and each user has $N_{R}=2$ antennas.
The path gains $g_{k,n}$'s are generated using the path loss model
(\textquotedblleft{}Urban Macro NLOS\textquotedblright{} model) in
\cite{3gpp_Rel9}. The size of each media file is $F_{l}=F_{0}=600$M
Bytes and the steaming rate is $\mu_{l}=\mu_{0},\forall l$. Assume
that each user independently accesses the $l$-th media file with
probability $\rho_{l}$, and we set $\boldsymbol{\rho}=\left[\rho_{1},...,\rho_{6}\right]=\left[0.6,0.3,0.08,0.01,0.005,0.005\right]$,
which represents the popularity of different media files. Note that
$\boldsymbol{\rho}$ is only used to generate the realizations of
URP $\pi$ and the control algorithms do not have the knowledge of
$\boldsymbol{\rho}$. We assume that the URP\textbf{ }$\pi$ changes
every 10000 time slots, i.e., $T_{i+1}-T_{i}=10000$. The other system
parameters are set as
\begin{eqnarray*}
B_{W}=1\textrm{MHz}, & \tau=5\textrm{ms}, & M=7.
\end{eqnarray*}

\subsection{Convergence of Algorithm SP and LC}

\begin{figure}
\begin{centering}
\includegraphics[width=85mm]{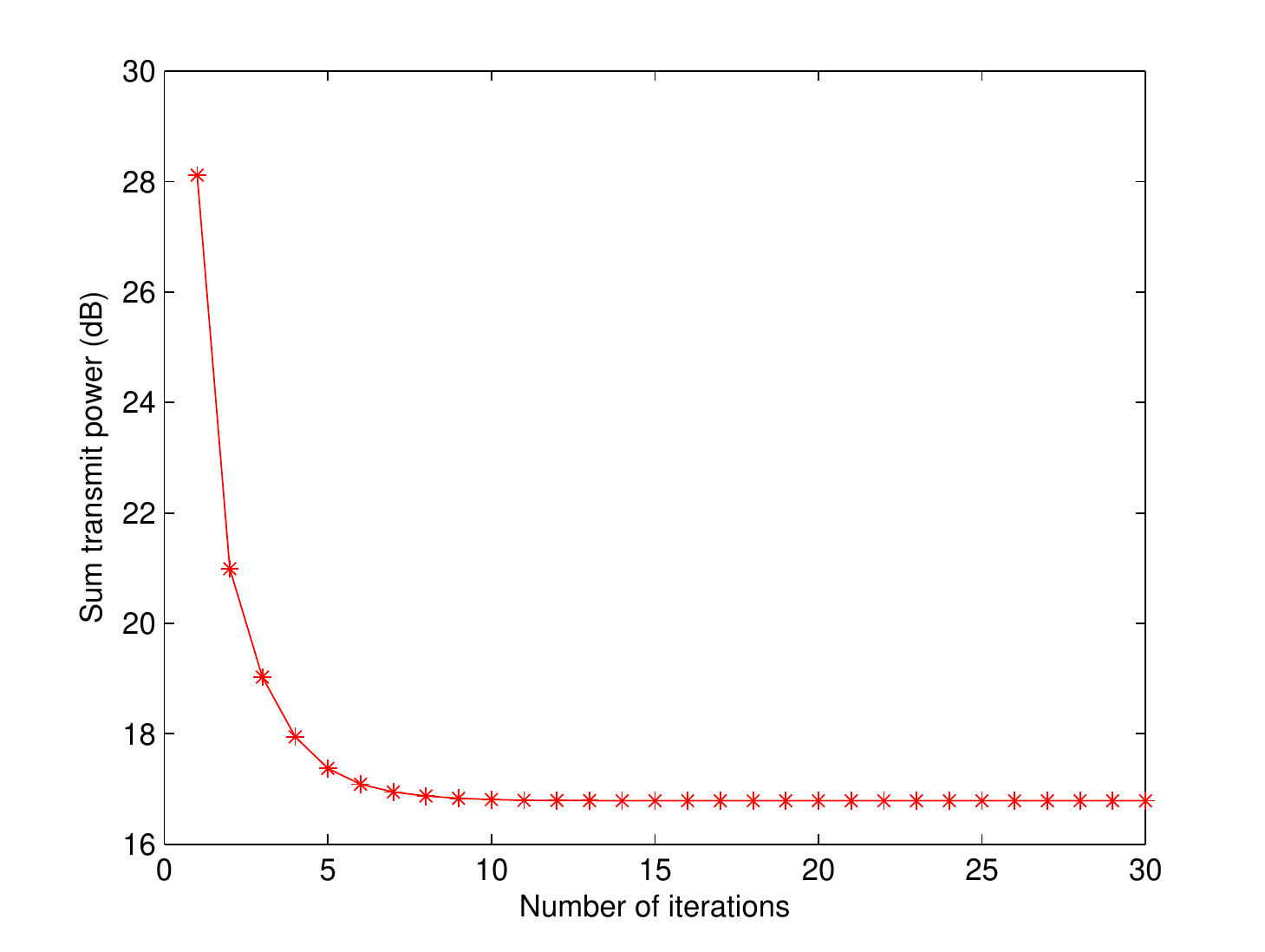}
\par\end{centering}

\caption{\label{fig:convSP}{\small{Objective value of $\mathcal{P}_{S}\left(\pi,\mathbf{H}\right)$
versus the number of iterations of Algorithm SP.}}}
\end{figure}

\begin{figure}
\begin{centering}
\includegraphics[width=85mm]{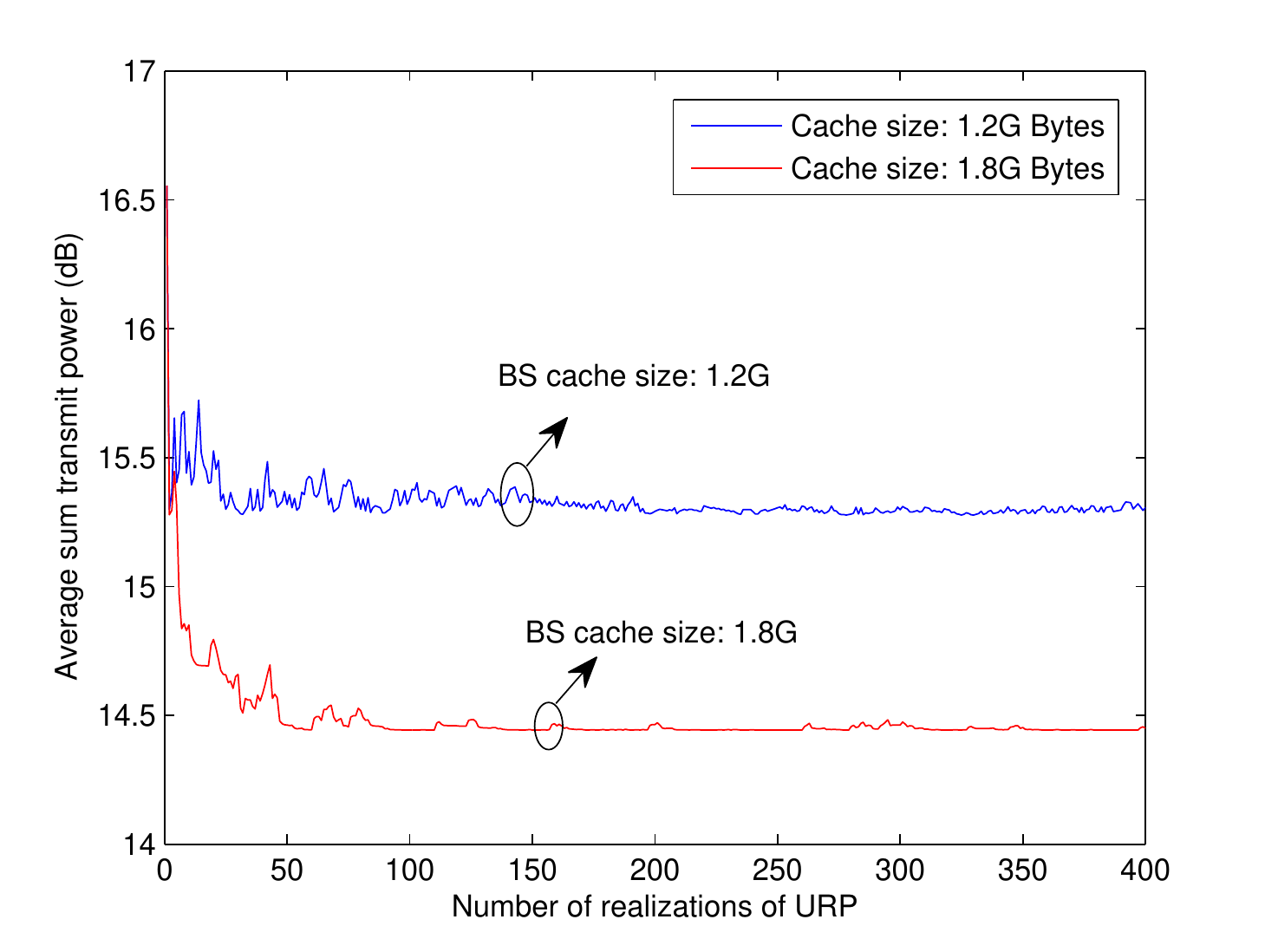}
\par\end{centering}

\caption{\label{fig:convLC}{\small{Objective value of }}$\mathcal{P}_{L}\left(\mathcal{V}^{*}\right)${\small{
versus the number of iterations of Algorithm LC.}}}
\end{figure}

Consider normal user placement and set the streaming rate as $\mu_{l}=\mu_{0}=2$Mbits/s,
$\forall l$. In Fig. \ref{fig:convSP}, we plot the objective value
$P\left(\mathbf{V}\right)$ of $\mathcal{P}_{S}\left(\pi,\mathbf{H}\right)$
versus the number of iterations for a single realization of $\pi,\mathbf{H}$.
In Fig. \ref{fig:convLC}, we plot the objective value $\psi\left(\mathbf{q},\mathcal{V}^{*}\right)$
of $\mathcal{P}_{L}\left(\mathcal{V}^{*}\right)$ versus the number
of realizations of $\pi$ (i.e., the number of iterations of Algorithm
LC) for different BS cache size $B_{C}$. It can be seen that both
Algorithm SP and LC quickly converge.

\subsection{Advantage of the Proposed Solution w.r.t. Baselines}

The following baselines are considered.

\textbf{Baseline 1 (Coordinated MIMO):} The BS has no cache. The physical
layer reduces to the coordinated MIMO precoding scheme in Section
\ref{sub:Cache-enabled-Opportunistic-CoMP}. At each time slot, the
precoding matrices are obtained by solving $\mathcal{P}_{S}\left(\pi,\mathbf{H}\right))$
using Algorithm SP.

\textbf{Baseline 2 (Conventional CoMP):} The BS has no cache. The
BSs employ CoMP transmission to serve the users by exchanging both
CSI and payload data via backhaul. At each time slot, the precoding
matrices are obtained by solving $\widetilde{\mathcal{P}}_{S}\left(\pi,\mathbf{H}\right))$
using Algorithm SP.

\textbf{Baseline 3: (Algorithm SP with Uniform Caching):} The long-term
cache control is given by a uniform caching scheme where $q_{l}=\frac{B_{C}}{LF_{0}},\forall l$.
The precoding matrices are the same as that of the proposed solution.

\begin{figure}
\begin{centering}
\includegraphics[width=85mm]{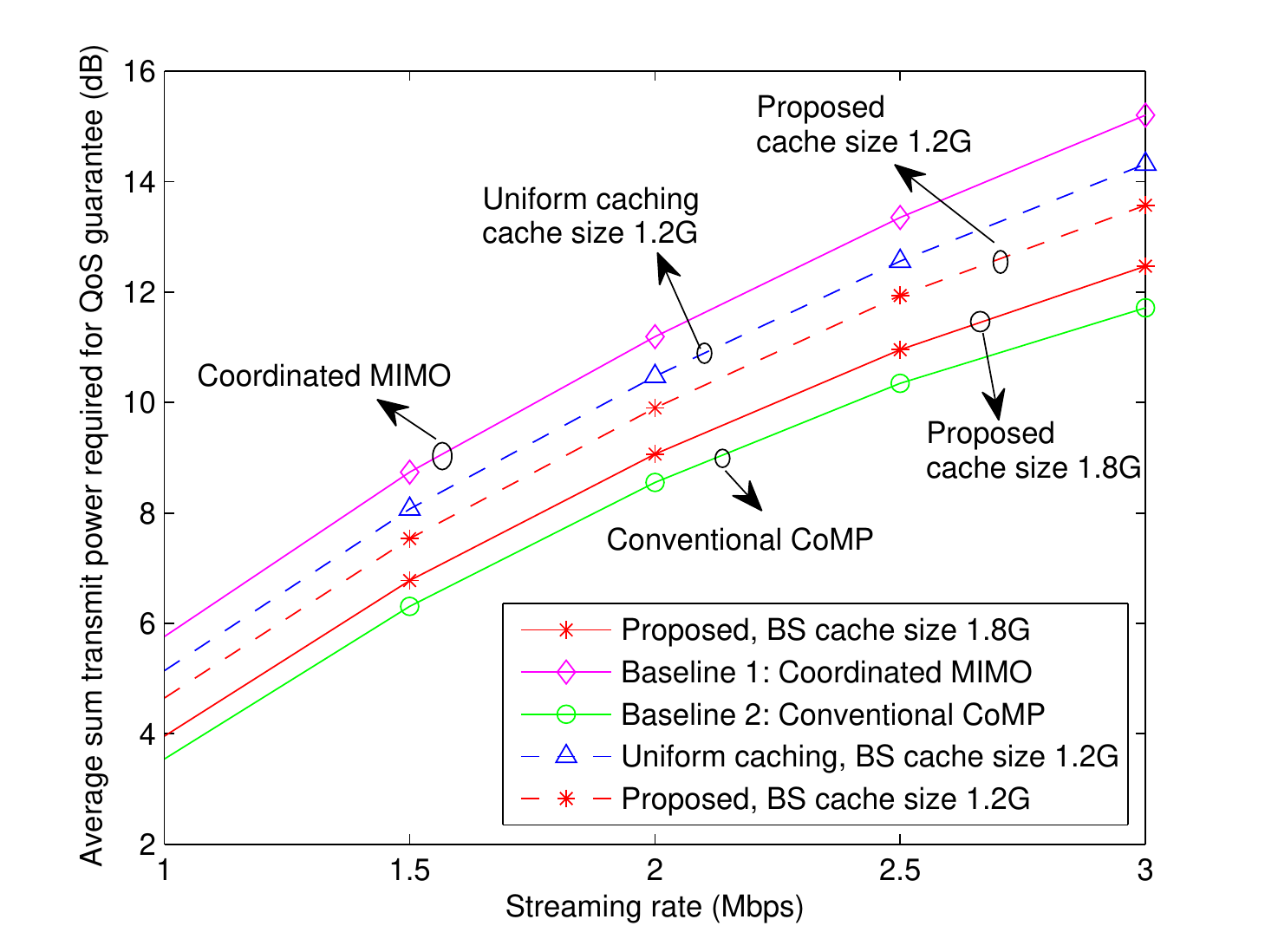}
\par\end{centering}

\caption{\label{fig:Pow_varmu0}{\small{Average sum transmit power required
for QoS guarantee versus the streaming rate for normal user placement.}}}
\end{figure}

\begin{figure}
\begin{centering}
\includegraphics[width=85mm]{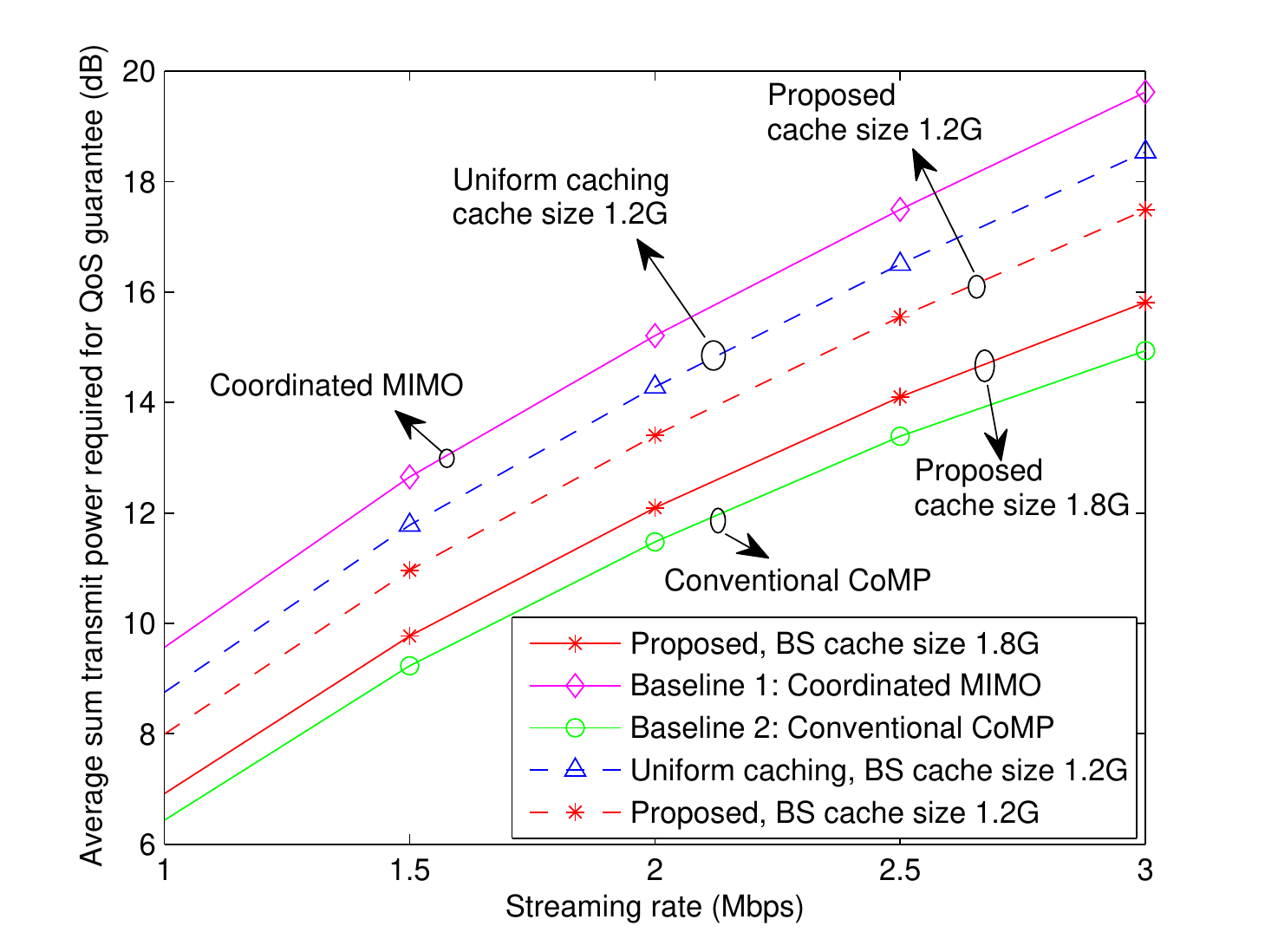}
\par\end{centering}

\caption{\label{fig:Pow_vargc}{\small{Average sum transmit power required
for QoS guarantee versus the streaming rate for edge user placement.}}}
\end{figure}

We first consider normal user placement. In Fig. \ref{fig:Pow_varmu0},
we plot the average sum transmit power required for no media playback
interruption versus the streaming rate $\mu_{0}$ for difference schemes.%
{} The proposed solution has a significant performance gain over baseline
1 (coordinated MIMO), and the gain increases as the BS cache size
increases. The proposed solution also has a large gain over baseline
3 (uniform caching) with the same BS cache size. This demonstrates
the advantage of the proposed MDS-coded random cache data structure.
As the BS cache size increases, the performance of the proposed solution
approaches that of baseline 2 (conventional CoMP). Specifically, when
the BS cache size $B_{C}=1.8$G Bytes, which is only about half of
the total size of all the media files, the performance is already
close to baseline 2. 

Then we consider edge user placement. In Fig. \ref{fig:Pow_vargc},
we plot the average sum transmit power required for no media playback
interruption versus the streaming rate $\mu_{0}$ for difference schemes.
Similar results as in Fig. \ref{fig:Pow_varmu0} can be observed.
Moreover, it can be seen that the performance gain of the proposed
solution w.r.t. coordinated MIMO is larger under the edge user placement.
This shows that a larger CoMP gain can be achieved when the cross
links are stronger.%
{} 

Finally, we show that the proposed solution has significant gain over
all the baselines even if we take into account the backhaul overhead.
Table \ref{tab:cacheupdateload} compares the average backhaul consumption
of different schemes when the streaming rate is 2Mbps and edge user
placement is adopted. It can be seen that the average backhaul consumption
of the proposed solution is much smaller than the conventional CoMP
and the coordinated MIMO. Moreover, the average backhaul consumption
decreases as the BS cache size $B_{C}$ increases. The backhaul consumption
under normal user placement is similar and is omitted for conciseness.

\section{Conclusion\label{sec:Conlusion} }

We propose a cache-induced opportunistic CoMP scheme for wireless
media streaming in MIMO interference networks. By caching a portion
of the media files at the BSs, the BSs are able to opportunistically
employ CoMP without expensive backhaul. We first propose a novel MDS-coded
random cache data structure which can significantly improve the CoMP
opportunities. We then formulate a mixed-timescale joint optimization
problem for MIMO precoding and cache control. The long-term cache
control is used to achieve the best tradeoff between CoMP opportunities
and BS cache size. The short-term MIMO precoding is to guarantee the
QoS requirements for given cache control. We propose a polynomial
complexity precoding algorithm to find a stationary point of the MIMO
precoding problem and a stochastic subgradient algorithm to find the
cache control solution. The proposed solution can achieve significant
performance gain over coordinated MIMO techniques with smaller backhaul
loading as demonstrated by numerical simulations.

\appendix

\subsection{Proof of Proposition \ref{prop:Optimal-number-ofd}\label{sub:Proof-of-Proposition-Optd} }

For convenience, let $\textrm{span}\left(\mathbf{A}\right)$ represent
the subspace spanned by the columns of a matrix $\mathbf{A}$ and
$\textrm{orth}\left(\mathbf{A}\right)$ represent a set of orthogonal
basis of $\textrm{span}\left(\mathbf{A}\right)$. For given precoding
matrices $\mathbf{V}^{'}=\left\{ \mathbf{V}_{m,k}^{'}\in\mathbb{C}^{N_{T}\times d_{m,k}}:\forall m,k\right\} $,
let $\overline{\mathbf{V}}_{m,k}^{'}=\textrm{orth}\left(\mathbf{H}_{m,k,k}^{\dagger}\right)\textrm{orth}\left(\mathbf{H}_{m,k,k}^{\dagger}\right)^{\dagger}\mathbf{V}_{m,k}^{'}$
denote a precoding matrix obtained by the projection of $\mathbf{V}_{m,k}^{'}$
on the subspace $\textrm{span}\left(\mathbf{H}_{m,k,k}^{\dagger}\right)$.
It can be verified that $\mathbf{H}_{m,k,k}\mathbf{V}_{m,k}^{'}=\mathbf{H}_{m,k,k}\overline{\mathbf{V}}_{m,k}^{'}$
and $\mathbf{\Omega}_{m,k}^{'}-\overline{\mathbf{\Omega}}_{m,k}^{'}\succeq\mathbf{0}$,
where $\mathbf{\Omega}_{m,k}^{'}$ and $\overline{\mathbf{\Omega}}_{m,k}^{'}$
are the interference-plus-noise covariance matrices respectively resulted
from the precoding matrices $\mathbf{V}^{'}$ and $\overline{\mathbf{V}}^{'}=\left\{ \overline{\mathbf{V}}_{m,k}^{'}\in\mathbb{C}^{N_{T}\times d_{m,k}}:\forall m,k\right\} $.
As a result, we have 
\begin{equation}
R_{k}\left(\mathbf{H},\overline{\mathbf{V}}^{'}\right)\geq R_{k}\left(\mathbf{H},\mathbf{V}^{'}\right),\: P_{k}\left(\overline{\mathbf{V}}^{'}\right)\leq P_{k}\left(\mathbf{V}^{'}\right).\label{eq:RatePowVp}
\end{equation}
Since $\textrm{Rank}\left(\overline{\mathbf{V}}_{m,k}^{'}\right)\leq\textrm{Rank}\left(\mathbf{H}_{m,k,k}\right)\leq d,\forall m,k$,
there exists $\mathbf{V}_{m,k}\in\mathbb{C}^{N_{T}\times d}$ such
that $\mathbf{V}_{m,k}\mathbf{V}_{m,k}^{\dagger}=\overline{\mathbf{V}}_{m,k}^{'}\overline{\mathbf{V}}_{m,k}^{'\dagger},\forall m,k$,
from which it follows that 
\begin{equation}
R_{k}\left(\mathbf{H},\mathbf{V}\right)=R_{k}\left(\mathbf{H},\overline{\mathbf{V}}^{'}\right),\: P_{k}\left(\mathbf{V}\right)=P_{k}\left(\overline{\mathbf{V}}^{'}\right),\label{eq:RatepowVVp}
\end{equation}
where $\mathbf{V}=\left\{ \mathbf{V}_{m,k}\in\mathbb{C}^{N_{T}\times d}:\forall m,k\right\} $.
Then Proposition \ref{prop:Optimal-number-ofd} follows from (\ref{eq:RatePowVp})
and (\ref{eq:RatepowVVp}).

\subsection{Proof of Proposition \ref{prop:Feasibility-of-P}\label{sub:Proof-of-Proposition-PF}}

Obviously, we only need to prove that there exists $\left\{ \mathbf{V}\left(\pi,\mathbf{H}\right):\forall\pi,\mathbf{H}\right\} $
such that $P\left(\mathbf{V}\left(\pi,\mathbf{H}\right)\right)<\infty,\:\mathbf{V}\left(\pi,\mathbf{H}\right)\in\mathcal{D}_{\mathbf{v}}\left(\pi,\mathbf{H}\right)$
with probability one and $\textrm{E}\left[P\left(\mathbf{V}\left(\pi,\mathbf{H}\right)\right)|\pi\right]$
is bounded for all $\pi$. Consider a simple precoding scheme where
$\mathbf{V}_{m,k}\left(\pi,\mathbf{H}\right)=\mathbf{0},\forall k\neq m$
and $\mathbf{V}_{m,k}\left(\pi,\mathbf{H}\right)=\sqrt{\frac{p_{k}}{d}}\left[\mathbf{e}_{m,k,k},\cdots,\mathbf{e}_{m,k,k}\right]\in\mathbb{C}^{N_{T}\times d},\forall k=m$,
where $\mathbf{e}_{m,k,k}$ is the dominate eigenvector of $\mathbf{H}_{m,k,k}$
and $p_{k}$ is chosen such that $\frac{B_{W}}{M}\textrm{log}\left(\mathbf{I}+\mathbf{H}_{k,k,k}\mathbf{V}_{k,k}\mathbf{V}_{k,k}^{\dagger}\mathbf{H}_{k,k,k}^{\dagger}\right)=\mu_{\pi_{k}}\ln2$.
Then if $\sum_{k=1}^{K}p_{k}<\infty$, we have $P\left(\mathbf{V}\left(\pi,\mathbf{H}\right)\right)=\sum_{k=1}^{K}p_{k}<\infty$
and $\mathbf{V}\left(\pi,\mathbf{H}\right)\in\mathcal{D}_{\mathbf{v}}\left(\pi,\mathbf{H}\right)$.
It can be verified that 
\begin{equation}
p_{k}\leq d\left[\textrm{Tr}\left(\mathbf{H}_{k,k,k}\mathbf{H}_{k,k,k}^{\dagger}\right)\right]^{-1}\left(e^{\frac{M\mu_{\pi_{k}}\ln2}{B_{W}}}-1\right).\label{eq:pbound}
\end{equation}
Since $\textrm{Pr}\left[\textrm{Tr}\left(\mathbf{H}_{k,k,k}\mathbf{H}_{k,k,k}^{\dagger}\right)=0\right]=0,\forall k$,
it follows that $\sum_{k=1}^{K}p_{k}<\infty$ with probability one.
Moreover, we have 
\begin{eqnarray*}
 &  & \textrm{E}\left[P\left(\mathbf{V}\left(\pi,\mathbf{H}\right)\right)|\pi\right]\\
 & \leq & d\sum_{k=1}^{K}\left(e^{\frac{M\mu_{\pi_{k}}\ln2}{B_{W}}}-1\right)\textrm{E}\left[\textrm{Tr}\left(\mathbf{H}_{k,k,k}\mathbf{H}_{k,k,k}^{\dagger}\right)^{-1}\right]\\
 & < & \infty,
\end{eqnarray*}
where the last inequality holds because $\textrm{Tr}\left(\mathbf{H}_{k,k,k}\mathbf{H}_{k,k,k}^{\dagger}\right)$
(after proper normalization) has a chi-squared distribution with $2N_{R}N_{T}\geq4$
degrees of freedom. This completes the proof.

\subsection{Proof of Theorem \ref{thm:Equivalence-between-PSandPMSE}\label{sub:Proof-of-TheoremEquiPS11}}

Theorem \ref{thm:Equivalence-between-PSandPMSE} can be proved by
contradiction. Suppose that $\mathbf{V}^{\star}$ is not the optimal
solution of $\mathcal{P}_{S}\left(\pi,\mathbf{H}\right)$. Then there
exists $\mathbf{V}\in\mathcal{D}_{\mathbf{v}}\left(\pi,\mathbf{H}\right)$
such that $P\left(\mathbf{V}\right)<P\left(\mathbf{V}^{\star}\right)$.
Let $\mathbf{U}=\left\{ \mathbf{U}_{m,k}:\forall m,k\right\} $ and
$\mathbf{W}=\left\{ \mathbf{\mathbf{W}}_{m,k}:\forall m,k\right\} $,
where $\mathbf{U}_{m,k}=\left(\mathbf{\Omega}_{m,k}+\mathbf{H}_{m,k,k}\mathbf{V}_{m,k}\mathbf{V}_{m,k}^{\dagger}\mathbf{H}_{m,k,k}^{\dagger}\right)^{-1}\mathbf{H}_{m,k,k}\mathbf{V}_{m,k}$
is the MMSE receiver corresponding to $\mathbf{V}$ and $\mathbf{\mathbf{W}}_{m,k}=\left(\mathbf{I}-\mathbf{U}_{m,k}^{\dagger}\mathbf{H}_{m,k,k}\mathbf{V}_{m,k}\right)^{-1}$.
Then it can be verified that%
{} $\left(\mathbf{W},\mathbf{U},\mathbf{V}\right)$ satisfies the MSE
constraint in Problem (\ref{eq:PMSE}), which contradicts with the
fact that $\left(\mathbf{W}^{\star},\mathbf{U}^{\star},\mathbf{V}^{\star}\right)$
is the optimal solution of Problem (\ref{eq:PMSE}). This completes
the proof.

\subsection{Proof of Theorem \ref{thm:Convergence-of-AlgPS}\label{sub:Proof-of-TheoremconvPS}}

Following similar analysis as in the proof of \cite[Theorem 3]{Luo_TSP11_WMMSE},
it can be shown that any limit point $\left(\mathbf{W}^{*},\mathbf{U}^{*},\mathbf{V}^{*}\right)$
is a stationary point of Problem (\ref{eq:PMSE}). As a result, $\mathbf{V}^{*}$
and the corresponding Lagrange multiplier vector $\boldsymbol{\lambda}^{*}$
satisfies the following KKT conditions 
\begin{eqnarray}
\nabla_{\mathbf{V}_{m,k}}L\left(\boldsymbol{\lambda}^{*},\mathbf{V}^{*},\mathbf{U}^{*},\mathbf{W}^{*}\right) & = & 0,\:\forall m,k.\nonumber \\
\lambda_{k}^{*}\left(\overline{\mu}_{\pi_{k}}-\overline{R}_{k}\left(\mathbf{H},\mathbf{V}^{*},\mathbf{U}^{*},\mathbf{W}^{*}\right)\right) & = & 0,\:\forall k,\label{eq:KKTPS}
\end{eqnarray}
and $\overline{R}_{k}\left(\mathbf{H},\mathbf{V}^{*},\mathbf{U}^{*},\mathbf{W}^{*}\right)\geq\overline{\mu}_{\pi_{k}}$,
where {\small{
\[
\overline{R}_{k}\left(\mathbf{H},\mathbf{V},\mathbf{U},\mathbf{W}\right)=d-\frac{\sum_{m=1}^{M}\left(\textrm{Tr}\left(\mathbf{\mathbf{W}}_{m,k}\mathbf{E}_{m,k}\right)-\textrm{log}\left|\mathbf{\mathbf{W}}_{m,k}\right|\right)}{M}
\]
}}Moreover, the step 1 and step 2 of Algorithm SP ensure that $\mathbf{U}_{m,k}^{*}$
is the MMSE receiver corresponding to $\mathbf{V}^{*}$ and $\mathbf{\mathbf{W}}_{m,k}^{*}=\left(\mathbf{I}-\mathbf{U}_{m,k}^{*\dagger}\mathbf{H}_{m,k,k}\mathbf{V}_{m,k}^{*}\right)^{-1}$,
from which it follows that 
\begin{equation}
R_{k}\left(\mathbf{H},\mathbf{V}^{*}\right)=\frac{B_{W}}{\ln2}\overline{R}_{k}\left(\mathbf{H},\mathbf{V}^{*},\mathbf{U}^{*},\mathbf{W}^{*}\right)\geq\mu_{\pi_{k}}.\label{eq:Rkbd}
\end{equation}
Using chain rule, it can be shown that (please refer to \cite[Appendix C]{Luo_TSP11_WMMSE}
for the detailed derivation)
\begin{equation}
\frac{B_{W}}{\ln2}\nabla_{\mathbf{V}_{m,k}}\overline{R}_{k}\left(\mathbf{H},\mathbf{V}^{*},\mathbf{U}^{*},\mathbf{W}^{*}\right)=\nabla_{\mathbf{V}_{m,k}}R_{k}\left(\mathbf{H},\mathbf{V}^{*}\right).\label{eq:DRbar}
\end{equation}
It follows from (\ref{eq:KKTPS}), (\ref{eq:Rkbd}) and (\ref{eq:DRbar})
that
\begin{eqnarray*}
\nabla_{\mathbf{V}_{m,k}}\mathcal{L}_{\pi,\mathbf{H}}\left(\frac{\ln2}{B_{W}}\boldsymbol{\lambda}^{*},\mathbf{V}^{*}\right) & = & 0,\:\forall m,k,\\
\frac{\ln2}{B_{W}}\lambda_{k}^{*}\left(\mu_{\pi_{k}}-R_{k}\left(\mathbf{H},\mathbf{V}^{*}\right)\right) & = & 0,\:\forall k,
\end{eqnarray*}
where $\mathcal{L}_{\pi,\mathbf{H}}\left(\boldsymbol{\lambda},\mathbf{V}\right)$
is the Lagrange function of $\mathcal{P}_{S}\left(\pi,\mathbf{H}\right)$:
\[
\mathcal{L}_{\pi,\mathbf{H}}\left(\boldsymbol{\lambda},\mathbf{V}\right)=P\left(\mathbf{V}\right)+\sum_{k=1}^{K}\lambda_{k}\left(\mu_{\pi_{k}}-R_{k}\left(\mathbf{H},\mathbf{V}\right)\right).
\]
Hence, $\mathbf{V}^{*}$ is a stationary point of $\mathcal{P}_{S}\left(\pi,\mathbf{H}\right)$
with $\frac{\ln2}{B_{W}}\boldsymbol{\lambda}^{*}$ as the corresponding
Lagrange multiplier vector.

\subsection{Proof of Lemma \ref{lem:Convexity-of-PL}\label{sub:Proof-of-PropositionCPL}}

The initial point given in (\ref{eq:initVta}) ensures that the initial
objective value of $\widetilde{\mathcal{P}}_{S}\left(\pi,\mathbf{H}\right)$
is equal to $P\left(\mathbf{V}^{*}\left(\pi,\mathbf{H}\right)\right)$.
In each iteration of Algorithm SP, the objective value is strictly
decreased before converging to a stationary point of $\widetilde{\mathcal{P}}_{S}\left(\pi,\mathbf{H}\right)$.
Hence, we must have $\widetilde{P}\left(\widetilde{\mathbf{V}}^{*}\left(\pi,\mathbf{H}\right)\right)\leq P\left(\mathbf{V}^{*}\left(\pi,\mathbf{H}\right)\right)$
and $\textrm{E}\left[\left.\widetilde{P}\left(\widetilde{\mathbf{V}}^{*}\left(\pi,\mathbf{H}\right)\right)\right|\pi\right]\leq\textrm{E}\left[\left.P\left(\mathbf{V}^{*}\left(\pi,\mathbf{H}\right)\right)\right|\pi\right]$.
It can be verified that $\textrm{min}_{k}\left\{ q_{\pi_{k}}\right\} $
is a concave function w.r.t. $\mathbf{q}$. Moreover, it follows from
$\textrm{E}\left[\left.\widetilde{P}\left(\widetilde{\mathbf{V}}^{*}\left(\pi,\mathbf{H}\right)\right)\right|\pi\right]\leq\textrm{E}\left[\left.P\left(\mathbf{V}^{*}\left(\pi,\mathbf{H}\right)\right)\right|\pi\right]$
that $\overline{P}_{\pi}\left(\mathbf{q},\mathcal{V}^{*}\right)$
is a decreasing linear function of $\textrm{min}_{k}\left\{ q_{\pi_{k}}\right\} $
for $q_{l}\in\left[0,1\right],\forall l$. Then using the vector composition
rule for convex function \cite{Boyd_04Book_Convex_optimization},
$\overline{P}_{\pi}\left(\mathbf{q},\mathcal{V}^{*}\right)$ is a
convex and non-differentiable function of $\mathbf{q}$ for fixed
$\mathcal{V}^{*}$, which implies that $\mathcal{P}_{L}\left(\mathcal{V}^{*}\right)$
is also a convex problem.

\subsection{Proof of Lemma \ref{lem:subconvcond}\label{sub:Proof-of-Lemmasubconv}}

Note that $\psi\left(\mathbf{q},\mathcal{V}^{*}\right)=\textrm{E}\left[\phi\left(\mathbf{q},\pi,\mathbf{H}\right)\right]$,
where 
\begin{eqnarray*}
\phi\left(\mathbf{q},\pi,\mathbf{H}\right) & = & \left(1-\textrm{min}_{k}\left\{ q_{\pi_{k}}\right\} \right)P\left(\mathbf{V}^{*}\left(\pi,\mathbf{H}\right)\right)\\
 &  & +\textrm{min}_{k}\left\{ q_{\pi_{k}}\right\} \widetilde{P}\left(\widetilde{\mathbf{V}}^{*}\left(\pi,\mathbf{H}\right)\right),
\end{eqnarray*}
is a convex function of $\mathbf{q}$. It is easy to see that for
each $\pi$ and $\mathbf{H}$, 
\begin{eqnarray*}
 &  & G\left(\mathbf{q},\pi,\mathbf{H}\right)\\
 & = & 1\left(l=\pi_{k^{*}}\right)\left[\widetilde{P}\left(\widetilde{\mathbf{V}}^{*}\left(\pi,\mathbf{H}\right)\right)-P\left(\mathbf{V}^{*}\left(\pi,\mathbf{H}\right)\right)\right]
\end{eqnarray*}
is a subgradient of $\phi\left(\mathbf{q},\pi,\mathbf{H}\right)$.
Hence, for each $\pi,\mathbf{H}$ and any $\mathbf{q}^{'}$, we have
\begin{equation}
\phi\left(\mathbf{q}^{'},\pi,\mathbf{H}\right)\geq\phi\left(\mathbf{q},\pi,\mathbf{H}\right)+G\left(\mathbf{q},\pi,\mathbf{H}\right)^{T}\left(\mathbf{q}^{'}-\mathbf{q}\right).\label{eq:faisub}
\end{equation}
Taking expectation for both sides of (\ref{eq:faisub}), we have
\[
\psi\left(\mathbf{q}^{'},\mathcal{V}^{*}\right)\geq\psi\left(\mathbf{q},\mathcal{V}^{*}\right)+\textrm{E}\left[G\left(\mathbf{q},\pi,\mathbf{H}\right)\right]^{T}\left(\mathbf{q}^{'}-\mathbf{q}\right),
\]
which implies that $\overline{G}\triangleq\textrm{E}\left[G\left(\mathbf{q},\pi,\mathbf{H}\right)\right]$
is a subgradient of $\psi\left(\mathbf{q},\mathcal{V}^{*}\right)$.
On the other hand, we have
\begin{eqnarray*}
\textrm{E}\left[\widehat{\frac{\partial\psi^{(i)}}{\partial q_{l}}}\right] & = & \textrm{E}\bigg[1\left(l=\pi_{k^{*}}\right)\bigg(\textrm{E}\left[\left.\widetilde{P}\left(\widetilde{\mathbf{V}}^{*}\left(\pi,\mathbf{H}\right)\right)\right|\pi\right]\\
 &  & -\textrm{E}\left[\left.P\left(\mathbf{V}^{*}\left(\pi,\mathbf{H}\right)\right)\right|\pi\right]\bigg)\bigg]=\overline{G}.
\end{eqnarray*}
Hence, $\widehat{\frac{\partial\psi^{(i)}}{\partial q_{l}}}$ is a
noisy unbiased subgradient of $\psi\left(\mathbf{q},\mathcal{V}^{*}\right)$.


%
\end{document}